\documentclass[fleqn,usenatbib]{mnras}

\usepackage{newtxtext,newtxmath}
\usepackage[T1]{fontenc}

\DeclareRobustCommand{\VAN}[3]{#2}
\let\VANthebibliography\thebibliography
\def\thebibliography{\DeclareRobustCommand{\VAN}[3]{##3}\VANthebibliography}

\usepackage{graphicx}	% Including figure files
\usepackage{amsmath}	% Advanced maths commands
\usepackage{amssymb}	% Extra maths symbols
\usepackage[table,xcdraw]{xcolor}  
\usepackage{afterpage}
\usepackage[normalem]{ulem} % allows to use \sout
\usepackage{placeins}

\usepackage{hyperref}
\DeclareRobustCommand{\VAN}[3]{#2} % set up for citation

\title[MOG in voids]{Fingerprints of modified gravity on galaxies in voids}

\author[P. Cataldi et al.]{
Pedro Cataldi,$^{1}$\thanks{Contact e-mail: pcataldi@iafe.uba.ar}
Susana Pedrosa,$^{1}$, Nelson Padilla ,$^{2}$ Susana Landau ,$^{3}$ Christian Arnold $^{4}$ and Baojiu Li $^{4}$
\\
$^{1}$Instituto de Astronom\'{\i}a y F\'{\i}sica del Espacio, CONICET-UBA, Casilla de Correos 67, Suc. 28, 1428, Buenos Aires, Argentina\\
$^{2}$Instituto de Astronom\'{\i}a Teórica y Experimental (IATE), UNC-CONICET, Laprida 854, X5000BGR, Córdoba, Argentina\\
$^{3}$Departamento de F\'{\i}sica, FCEN-UBA and IFIBA,  Av. Intendente Cantilo S/N 1428  Ciudad Autónoma de Buenos Aires, Argentina\\
$^{4}$Institute for Computational Cosmology, Department of Physics, Durham University, South Road, Durham DH1 3LE, UK}
\date{Accepted 2022 July 25. Received 2022 July 25; in original form 2022 June 16}

\pubyear{2022}

\begin{document}
\label{firstpage}
\pagerange{\pageref{firstpage}--\pageref{lastpage}}
\maketitle

% Abstract of the paper
\begin{abstract}
We search for detectable signatures of $f(R)$ gravity and its chameleon screening mechanism in the baryonic and dark matter (DM) properties of simulated void galaxies. The enhancement of the gravitational acceleration can have a meaningful impact on the scaling relations as well as on the halo morphology. The galaxy rotational velocity field (calculated with the velocity of the gas disc and the acceleration fields) deviates from the typical values of the Tully-Fisher Relation (TFR) in GR. For a given stellar mass,  $f(R)$ gravity tends to produce greater maximum velocities. On the other hand, the mass in haloes in $f(R)$ gravity is more concentrated than their counterparts in GR. This trend changes when the concentration is calculated with the dynamical density profile, which takes into account the unscreened outer regions of the halo. Stellar discs interact with the overall potential well in the central regions, modifying the morphology of the screening regions and reshaping them. We find a trend for galaxies with a more dominant stellar disc to deviate further from round screening regions. We find that  small haloes are less triaxial and more round in $f(R)$ than their GR counterparts. The difference between halo morphology becomes smaller in $f(R)$ haloes whose inner regions are screened. These results suggest possible observables that could unveil modified gravity effects on galaxies in voids in future cosmological tests of gravity. 
\end{abstract}

% Select between one and six entries from the list of approved keywords.
% Don't make up new ones.

%%%%%%%%%%%%%%%%%%%%%%%%%%%%%%%%%%%%%%%%%%%%%%%%%%

%%%%%%%%%%%%%%%%% BODY OF PAPER %%%%%%%%%%%%%%%%%%

\section{Introduction}
\label{sec:intro}

The discovery of the late time accelerated expansion of the Universe resulted in a challenging problem for theoretical physics, namely, the explanation of the physical mechanism that triggers this phenomenon. The solution offered by the standard cosmological model, i.e. the addition of a cosmological constant in Einstein's equations, has some theoretical problems, the most important one being the difficulty to explain its observed value. As a consequence, alternative cosmological models started to be considered, among them, many incorporate alternative theories to General Relativity (GR) to describe the gravitational interaction \citep[see for example the review of ][]{Clifton2012}, called modified gravity models (MOG). Another motivation for considering alternative theories of gravity to cosmological models is the Hubble tension, namely, the discrepancy in the value of the Hubble constant obtained with model independent supernovae observations \citep{Riess2021} with  the one inferred from the cosmic microwave background (CMB) data assuming a standard cosmological model \citep{Plank2014}.

A particular class of alternative  theories of gravity is $f(R)$ gravity in which the Ricci scalar $R$ in the Einstein-Hilbert action is replaced by a scalar function of $R$ \citep{DeFelice2010}. Although, it has been shown that $f(R)$ models do not alleviate the Hubble tension \citep[e.g.][]{ Odintsov2021}, these theories can be reformulated in terms of scalar-tensor theories with a coupling of the dynamical scalar field to matter which enhances the gravitational force. Nonrelativistic matter, such as dust, stars and gas, will feel this additional force, which in general leads to larger dynamically inferred masses. This discrepancy can be up to a factor of $\it 1/3$. Therefore, several of these gravity theories can be ruled out by local gravity tests such as fifth force experiments or solar system tests among others \citep{DeFelice2010}. However, certain variants of this model, known in the literature as chameleon $f(R)$ theories, can surpass this limitation due to the so called chameleon screening \citep{Brax2004}, which can suppress the fifth force in high density environments such as stars and galaxies \citep{Brax2008}. 

Regarding stars, many authors have discussed various observational consequences. For example, \citet{Davis2012} studied the effects of chameleon models upon the structure of the main sequences, pointing out that unscreened stars can be significantly more luminous. They also analyzed the effect of MOG on galactic luminosity in dwarf galaxies. Low mass stellar objects, such as red and brown dwarf stars, are excellent probes of these kind of theories. As \citet{Sakstein2015} has claimed, the radius of a brown dwarf, theoretically, can differ significantly from the GR prediction and upcoming surveys could potentially place new constraints.

Among galaxies, studies suggest that the fifth force effects must be screened within the Milky Way \citep[e.g.][]{Burrage2018, Sakstein2020} so that any viable $f(R)$ model is likely to have no detectable signature in our solar system. On the other hand, dwarf galaxies in low-density environments may remain unscreened. This kind of galaxies, in such environments, may exhibit manifestations of enhanced gravity in their internal dynamics and condensation of gas and stars. Therefore, dwarf galaxies are ideal scenarios to test the effect of MOG theories, in particular, $f(R)$ theories.

The effects of MOG may be difficult to disentangle from those of other astrophysical processes. To address this issue, most of the previous studies \citep[e.g.][]{Vikram2018,Vikram2018_shape} create a control sample of screened galaxies which are not expected to show any of the expected MOG effects. The division of the observed galaxies intro screened and unscreened catalogues is accomplished based on an estimate of the local value of the external and internal gravitational potential, using the methodology proposed by \citet{Cabre2012}.  

\citet{Jain2011} pointed out that for $f(R)$ gravity in galaxies, the fifth force affects the dark matter and HI gas disc but not the stellar disc due to the self-screening of stars, as being compact objects and hence have zero scalar charge. This means their motion in MOG is identical to that predicted by GR. Conversely, diffuse gas is unscreened and feels the full fifth-force present due to the modifications. This means that at fixed radius, the gaseous component of an unscreened galaxy should rotate with a higher velocity than the stellar component. This may lead to a separation of the stellar disc from the centre of mass of the dark matter and from the HI disc and result in observable distortions of the morphology and dynamics of the stellar disc \citep[e.g.][]{Vikram2018}.

\citet{Vikram2018_shape} focused on late-type dwarf galaxies and claimed that these are the most likely to be unscreened. \citet{Vikram2018} and \citet{Naik2019} compared the theoretical differences between the gaseous and stellar components of isolated dwarf galaxies rotational curves with the observational values obtained from VLT-FORS2 and SPARC samples. In this way, assuming Navarro-Frenk-White (NFW) \citep{Navarro1997} dark matter haloes, they were able to rule out values of $\mathrm{\left|f_{R0} \right|>10^{-6.0}}$ and $\mathrm{\left|f_{R0} \right|>10^{-6.5}}$, respectively.

Another important observational effect of MOG is the warping of the stellar disc. As the host dark matter halo of the galaxy moves along an external force, it pulls at the lagging stellar component. This external potential gradient when aligned with the axis of rotation of the stellar disc will warp the stellar disc in U-shaped form. This warp is expected to align with this potential gradient. \citet{Jain2011} estimated the warp to be of order 0.1 kpc. 

Regarding these two important features (offsets between stars and gas, and warping of the stellar disc), \citet{Desmond2020} used morphological indicators in galaxies to constrain the strength and range of the fifth force. They analyzed the $f(R)$ \citet{HuSawicki2007} model with $n=1$, superimposing analytical expressions using GR-based mock catalogues and found that for a background scalar field value $\mathrm{\left|f_{R0} \right|<1.4 \times 10^{-8.0}}$, all astrophysical objects are screened. Taking a different approach, we expect a similar analysis, but with MOG based simulations, may lead to different constraints different constraints for the background scalar field.  

Semi-analytical galaxy formation models combined with $f(R)$ gravity have demonstrated that the MOG effects on basic properties such as galaxy stellar mass functions and cosmic star formation rate densities are rather small and comparable to the uncertainties of the semi-analytical models (see for example the reviews of \citet{Llinares2018} and \citet{Vogelsberger2020}.)

Using a semi-analytical model, \citet{Naik2020} simulated satellites with a range of masses and orbits, together with a variety of strengths of the fifth force. The ratio of the cumulative number function of stars in the leading and trailing stream as a function of longitude from the satellite is computable from simulations, measurable from the stellar data and provided a direct test and constraint of chameleon gravity at the level of $\mathrm{\left|f_{R0} \right|=10^{-7.0}}$.

Fully self-consistent simulation studies of galaxy formation in such screened MOG models have only started very recently \citep[e.g.][]{Arnold2019}. Simulations so far have not explicitly implemented the effects that MOG has on stellar properties and the difficult task to discriminate the screening effects between stellar, dark matter and gas particles. \citet{Arnold2019}, using the fully hydrodinamical SHIBONE (Simulating Hydrodynamics Beyond Einstein) suite simulation, found that the enhancement of the halo mass function due to $f(R)$-gravity and its suppression due to feedback effects can be estimated from independent GR-hydro and $f(R)$ dark matter only simulations.  Low mass haloes are nevertheless more likely to be populated by galaxies in $f(R)$-gravity. 

In this paper we will consider deviations from GR exhibited in  numerical simulations of  $f(R)$ cosmology at galactic and group scales and study the effects of chameleon screening on baryonic physics. 

This paper is organised as follows. We review the theoretical models and numerical simulations used in our study in section \ref{sec:teo} and \ref{sec:simu}. In section \ref{sec:Galaxy_selection} we describe  our catalog of haloes in voids  for each cosmology run. In section \ref{sec:Results} we investigate the galaxy and halo properties, such as the scaling relations, galaxy morphology, halo concentration and the shape of the screening regions and the dark matter halo. We contrast our findings with the GR run to put in evidence the effects on the modified gravity. We summarise our main results in section \ref{sec:conclusions}.

\section{Theoretical models}
\label{sec:teo}
\subsection{\textit{f(R)}-Gravity}

Using the same framework as Einstein’s general relativity,  \textit{f(R)}-gravity introduces an additional scalar degree of freedom which leads to a fifth force, enhancing gravity by 4/3 in low density environments. This is achieved introducing a scalar function \textit{f(R)} of the Ricci scalar $R$ to the action by, 
\begin{equation}
    \mathrm{S=\int d^{4}x\sqrt{-g}\left [ \frac{R+f(R)}{16 \pi G} + \mathfrak{L}_{m} \right ]},
\end{equation}
where $g$ is the determinant of the metric $\mathrm{g_{\mu\nu}}$ and $\mathfrak{L}_{m}$ is the Lagrangian density of the matter fields.

Varying the action which respect to the metric leads to the field equation of \textit{f(R)}-gravity,
\begin{equation}
    \mathrm{G_{\mu \nu}+f_{R}R_{\mu \nu}-\left (\frac{f}{2}-\square f_{R} \right )g_{\mu \nu}-\triangledown _{\mu} \triangledown _{\nu}f_{R}=8\pi GT_{\mu \nu}} \; ,
\end{equation}
where $\mathrm{G_{\mu \nu }}$ and $\mathrm{R_{\mu \nu}}$ denote the components of the Einstein and Ricci tensor, respectively. The scalar degree of freedom, $\mathrm{f_{R}}$, is the derivative of the scalar function, $\mathrm{f_{R}\equiv df(R)/dR}$. The energy momentum tensor is $\mathrm{T_{\mu\nu}}$; covariant derivatives are written as $\mathrm{\triangledown _{\nu}}$ and $\mathrm{\square \equiv \triangledown_{\nu}\triangledown^{\nu}}$, where Einstein summation convention is used. 

As regards the viability of the $f(R)$ models, it should be stressed that they should behave very similar to the background expansion rate of the $\Lambda$CDM model, are stable to cosmological perturbations and avoid ghost states among many others \citep{HuSawicki2007,DeFelice2010}. Also, as commented above, in order to satisfy the constraints from local gravity tests, any successful $f(R)$ model should exhibit a chameleon screening mechanism, i.e., the equivalent scalar-tensor theory should be a chameleon field theory.

\subsection{Hu \& Sawicki model}

The Hu \& Sawicki model \citep{HuSawicki2007} is one of the most widely-studied models of modified gravity. One of the reasons for this is that the model is demonstrated to be able to be compatible with local gravity tests due to the chameleon effect. 

For this model, the proposed $f(R)$ function can be expressed as follows,
\begin{equation}
   \mathrm{f(R)=-m^{2}\frac{c_{1}\left ( \frac{R}{m^{2}} \right )^{n}}{c_{2} \left (\frac{R}{m^{2}} \right )  ^{n}+1}},
\end{equation}
where $c_1$, $c_2$, $n$ are dimensionless constants. We choose $n = 1$. $m$ is defined as,
\begin{equation}
\mathrm{m^2 = \frac{1}{(8315 \, \rm{Mpc})^2} \frac{\Omega_m h^2}{0.13}},   
\label{eq:m}
\end{equation}
Also, at large curvature with respect to $\mathrm{m^2}$,
\begin{equation}
 \mathrm{f(R) \simeq - \frac{c_1}{c_2} m^2 + \frac{c_1}{c_2^2} m^2 \left (\frac{m^2}{R} \right )^n}
 \label{eq:fRcurv}
\end{equation}
Moreover, any successful cosmological model must  describe the current accelerated expansion of the Universe. For this,  the following condition has to be satisfied when  $\mathrm{R \gg m^2}$,
\begin{equation}
\mathrm{f(R) \simeq - 2 \Lambda}  ,
\label{eq:cosmo}
\end{equation}
where $\Lambda$ is an effective cosmological constant. 

In this way, eqs. (\ref{eq:fRcurv}) and (\ref{eq:cosmo}) result in the following condition for the free parameters of the model,
\begin{equation}
\mathrm{\frac{c_1 m^2}{2 c_2} = \Lambda =  3 \frac{H_0^2}{c^2}\left(1-\Omega_m \right)},  
\label{eq:c1c2}
\end{equation}
where $\Omega_m$ is the total mass density parameter in the standard $\Lambda$CDM cosmological model. 

Thus, by setting the background value of the scalar field $\mathrm{f_{R0}=\frac{df}{dR}|_{R=R_0}}$ where $R_0$ is the current value of the Ricci scalar together with eq. (\ref{eq:c1c2}) and (\ref{eq:m}), all parameters of the model are determined give a fixed value of $n$. We define F6 and F5 as $\mathrm{\left|f_{R0} \right|=10^{-6.0}}$ and $\mathrm{\left|f_{R0} \right|=10^{-5.0}}$, respectively. 

In such theories the structure formation is governed by the following two equations, 
\begin{equation}
\begin{cases}
 \mathrm{\bigtriangledown ^{2}\Phi =\frac{16\pi G}{3}a^{2}\delta \rho -\frac{a^{2}}{6}\delta R(f_{R})},&  \\ 
 \mathrm{\bigtriangledown ^{2}f_{R} =-\frac{a^{2}}{3}\left [ \delta R(f_{R})+8\pi G\delta\rho  \right ]}&   
\end{cases}
\label{eq:PoissonMOG}
\end{equation} 
where $\Phi$ denotes the gravitational potential, $\rho$ de matter density and $\mathrm{\delta f_{R}=f_{R}(R)-f_{R}(\bar{R})}$, $\mathrm{\delta R= R - \bar{R}}$, $\mathrm{\delta \rho = \rho - \bar{\rho}}$ and the quantities with the overbar take the background values. The two coupled Poisson-like equations are more difficult to solve than the simple Poisson equations in GR, which are linear (i.e: $\mathrm{\bigtriangledown  ^{2}\Phi = 4 \pi G a^{2}\delta \rho}$). 

\subsection{The fifth force}

As we have described previously, the \citet{HuSawicki2007} $f(R)$  model is able to evade the stringent constrains of local gravity tests and still leave detectable signatures on large scales, making it an excellent model to explore the deviations from GR. 

Now, let us briefly recall the formulation of the  \citet{HuSawicki2007} $f(R)$ model in terms of a scalar-tensor theory. For this, first we define a chameleon field $\mathrm{\phi}$ as follows,
\begin{equation}
    \mathrm{e^{-\frac{2\beta \phi }{ M_{pl}}}=f_{R}+1}
\end{equation}
with $\mathrm{\beta=\sqrt{1/6}}$. Next, we apply the conformal transformation
\begin{equation}
    \mathrm{\tilde{g}_{\mu \nu} = e^{-\frac{2\beta \phi }{M_{pl}}}g_{\mu \nu}}
\end{equation}
In such way, the action can be expressed as,
\begin{equation}
   \mathrm{ S=\int d^{4}x\sqrt{-\tilde{g}}\left [ \frac{M_{pl}^2}{2}\tilde{R}-\frac{1}{2}\tilde{g}^{\mu \nu}\triangledown _{\nu} \phi \triangledown _{\mu} \phi -V(\phi) + \mathfrak{\tilde{L}} _{m}     \right ]},
\end{equation}where 
\begin{equation}
    \mathrm{V(\phi)=\frac{M_{pl}^2[Rf_{R}-f(R)]}{2(f_{R}+1)^2}}
\end{equation}
and $\tilde{R}$ is the Ricci scalar corresponding to the metric $\widetilde{g}_{\mu \nu}$. In the Newtonian limit, the field equations for $\mathrm{\phi}$ can be written as, 
\begin{equation}
   \mathrm{ \triangledown ^2 \phi = \frac{\partial V}{\partial \phi}+ \frac{\beta \rho}{M_{pl}} = \frac{dV_{\it eff}}{d\phi}}.
\label{eq:fieldchameleon}    
\end{equation}
For simplicity, we restrict our analysis to a spherically symmetric body of radius $R_{c}$. If the object is at least partially screened, the effective potential $\mathrm{V_{\it eff}}$ will reach its minimum inside the object in a so called 'screening' radius, $\mathrm{r_{_{s}}}$. The following condition is satisfied then
\begin{equation}
    \mathrm{\frac{\partial V}{ \partial \phi} = -\frac{\beta \rho}{M_{pl}}}.
\end{equation}
In this way, for $\mathrm{r<r_{_{s}}}$, $\mathrm{\phi = \phi_c =  constant}$. Far outside the sphere (for $\mathrm{r \gg R_{c} \ge r_{_{s}}}$) the field $\mathrm{\phi_{0}}$ is  given by the background value $f_{R0}$ of the scalar degree of freedom. In the region in between, one can linearise eq. (\ref{eq:fieldchameleon}) around the background value $\mathrm{\delta  \phi = \phi - \phi_{0}}$, 
\begin{equation}
    \mathrm{\triangledown ^2 \delta \phi = \frac{\partial^2 V}{\partial \phi ^2}\delta \phi + \frac{\beta \delta \rho}{M_{pl}}}
\end{equation}
If we integrate this equation twice and resubstitute the Newtonian potential for a spherical overdensity $\mathrm{d\phi_{N}/dr=GM(<r)/r^2}$, we arrive at an expression of the fifth force for $r>r_{s}$ \citep{Davis2012}. 
\begin{equation}
    \mathrm{F_{_{\mathrm{MOG}}}=\alpha \frac{GM(< r )}{r^2}\left [ 1 - \frac{M(r_{s})}{M(<r)} \right ]},
\label{eq:fifhtforce}    
\end{equation}
where $\alpha=2\beta ^2=1/3$ is the coupling strength of $f(R)$ gravity. We can estimate the screening radius $r_{s}$ as given by the integral equation \citep{Sakstein2013},
\begin{equation}
    \mathrm{\frac{\phi_{0}}{2\beta M_{pl}}=4\pi G \int_{r_{s}}^{R}r\rho(r)dr}.
\label{eq:integrars}    
\end{equation}
\subsection{Navarro-Frenk-White profile and the screening radius}

Finally, we assume that the density of the halo is given by a NFW-profile \citep{Navarro1997}
\begin{equation}
    \mathrm{\rho (r)= \frac{\rho_{c}}{(\frac{r}{r_{_{NFW}}})(1+\frac{r}{r_{_{NFW}}})^2}}
\label{eq:NFW}
\end{equation}
where $r_{_{NFW}}$ describes is the scale at which the profile slope is equal to 2 
and $\rho _{c}$ represents a characteristic density at the radius $\mathrm{r=r_{_{NFW}}}$. We define the virial mass $\mathrm{M_{200}}$, as the mass within the virial radius, $\mathrm{{200}}$, identified as the radius which encloses a density equal to $\sim$ 200 times the critical density. If we take account that 
\begin{equation}
  \mathrm{\frac{\phi_{0}}{2\beta M_{pl}}=-\frac{3}{2}\ln(f_{R0}+1)}  
\end{equation}
and when we insert this profile in the integral of eq. (\ref{eq:integrars}) the following equation is obtained,
\begin{equation}
   \mathrm{ -\frac{3}{2}\ln(f_{R0}+1)  =\frac{\phi_{0}}{2\beta M_{pl}}=\frac{4 \pi G \rho _{c} }{r_{NFW}}\int_{r_{s}^{{NFW}}}^{r_{200}}\frac{dr}{(1+\frac{r}{r_{NFW}})^{2}}}.
\end{equation}
This integral results in an expression for the screening radius $\mathrm{r_{s}^{{NFW}}}$,
\begin{equation}
    \mathrm{r_{_{s}}^{{NFW}}=\frac{r_{_{NFW}}}{\frac{1}{1+r_{200}/r_{_{NFW}}}-\frac{3ln(f_{R0}+1)}{8 \pi G \rho _{c}r^2_{_{NFW}}}}-r_{_{NFW}}}.
\label{eq:rsNFW}
\end{equation}
We defined haloes whose screening radius values are $\mathrm{0<r_{_{s}}< r_{200}}$, as partially screened haloes (from now on, \textit{PSH}). If the halo has not a screening region, we call it completely unscreened halo. 

\subsection{Motivation}
\label{sec:mot}

Until recently, a numerical study that can relate baryonic physics and MOG cosmology was not possible due to the absence of an efficient numerical code that could solve simultaneously, the modified Poisson equations (eqs. (\ref{eq:PoissonMOG})) and the hydrodynamic baryonic equations. 

In the case of $f(R)$ cosmology, many attempts were made to observationally constrain the strength of the scalar field \citep[e.g][]{Vikram2018_shape, Desmond2020} based on the baryon dynamics. With the introduction of efficient hydrodynamical cosmological numerical codes with MOG  (SHIBONE, \citep[][]{Arnold2019}), the study of the effects of MOG galaxy formation, galaxy morphology or scaling relations, in a numerical context, is now possible.

An interesting effect of \textit{PSH}, that constitutes the focus of our study, is that the morphology of the screened region seems to depend on the baryonic stellar disc density and the resulting modification of the gravity potential wells in the inner regions of the halo. A qualitative description of this phenomenon has been first reported by \citet{Naik2018}. If we take this effect into account, the popular parametrization of the screened region as a screened radius, assuming spherical shape of the screened region, should be taken as a first order approximation.

\begin{figure*}
\centering
\includegraphics[width=1.0\textwidth]{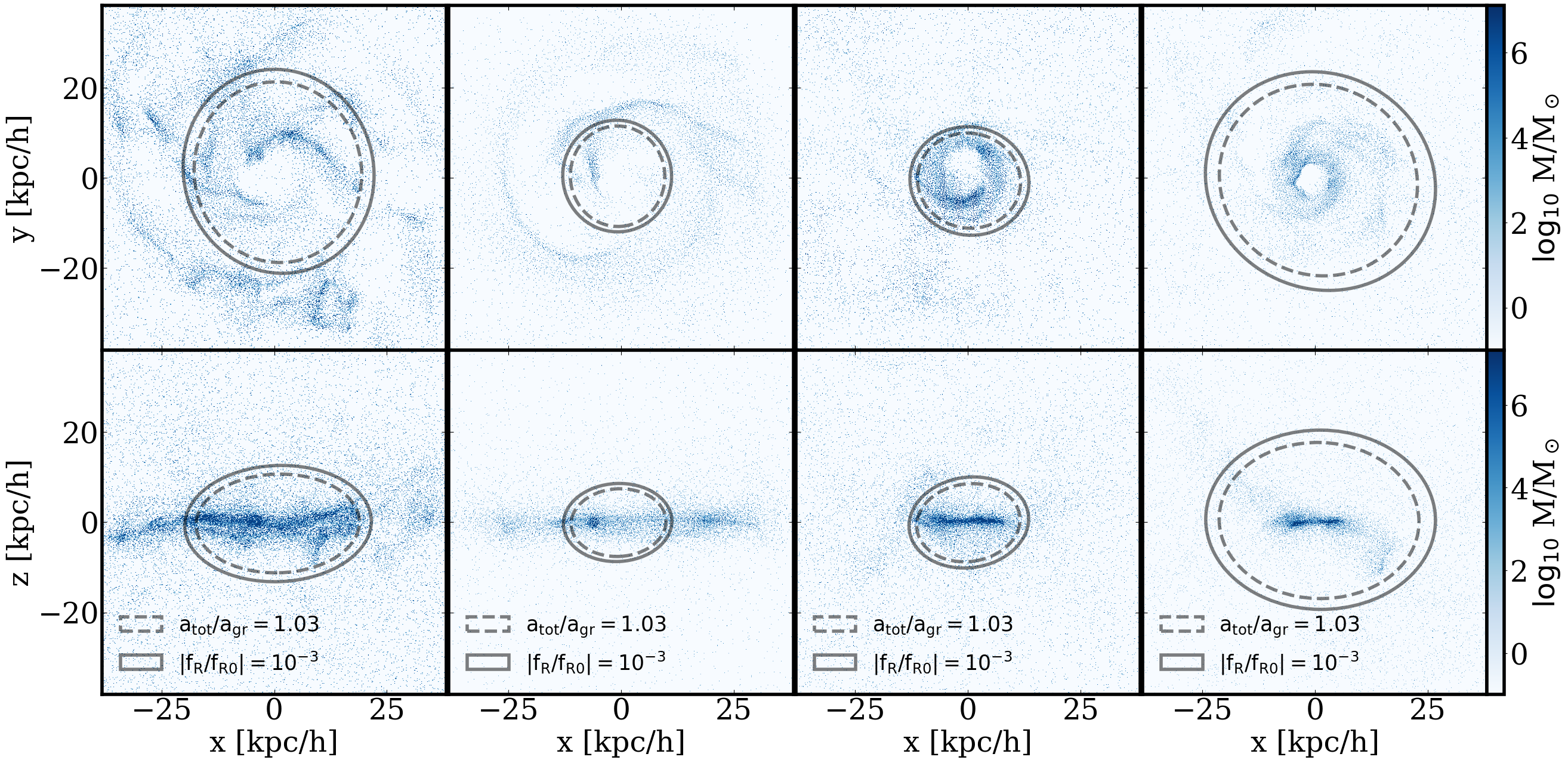}
\caption{(Colour Online) Face-on (top) and edge-on (bottom) galaxy contour maps of the scalar field ${f_{R}}$, for four of the chosen partially screen haloes (PSH). The contours (solid and dashed lines) shows the location of the screening surface for two different criteria: $\left |f_{R}/f_{R0}  \right | = 10^{-2})$ and $a_{\it tot}/a_{\it GR} = 1.03$. We chose to superimposed the figure on maps of the projected gas density, with a colour code of density in terms of $M_{\odot }$.}
\label{fig:1}
\end{figure*}

We aim to study (see Figure \ref{fig:1}), how disc galaxies can reshape this region and how to parametrize it. The extent of the screened regions depends on the chosen criteria ($\mathrm{\left |f_{R}/f_{R0}  \right |} = 10^{-2}$ and $\mathrm{a_{\it tot}/a_{gr}} = 1.03$). The morphology of the screened regions changes according to the gas density and, in particular, with the shape of the stellar disc frame. In the edge-on galaxy frame, the screened region has elliptical shape, while the face-on frame shows rounder shapes.

The motivation of this work is to find possible fingerprints of MOG effects on galaxies in underdense regions, where the fifth force is present. This can be possible with the comparison between simulation boxes with same initial conditions but with different cosmologies. We acknowledge the fact that astrophysical effects are relevant when we want to compare galaxy formation and the influence of cosmology. Different formation histories affect the resulting haloes, making it impossible to make a halo to halo comparison between different cosmologies. For this reason, we lean to a statistical approach searching for general trends.

\section{Numerical simulations}
\label{sec:simu}

\begin{table*}
\centering
\begin{tabular}{lllllll}
\hline
Simulation               & Hydro Model   & Cosmologies          & $N_{DM}$  & $N_{gas}$         & $m_{DM}[h^{-1}M_{\odot }]$ & $\bar{m}_{gas}[h^{-1}M_{\odot }]$ \\ \hline
Full-physics, 62 Mpc box & TNG-model     & $\Lambda$CDM, F6, F5 & $512^{3}$ & $\approx 512^{3}$ & $1.3 \times 10^{8}$        & $\approx 3.1 \times 10^{7}$       \\
\textbf{Full-physics, 25 Mpc box} & \textbf{TNG-model}     & \textbf{$\mathbf{\Lambda}$CDM, F6, F5} & $\mathbf{512^{3}}$ & $\mathbf{\approx 512^{3}}$ & $\mathbf{8.4 \times 10^{6}}$ & $\mathbf{\approx 2.2\times 10^{6}}$  \\
Non-rad, 62 Mpc box      & Non-radiative & $\Lambda$CDM, F6, F5 & $512^{3}$ & $\approx 512^{3}$ & $1.3 \times 10^{8}$        & $\approx 3.6\times 10^{7}$        \\
DM-only, 62 Mpc box      & -             & $\Lambda$CDM, F6, F4 & $512^{3}$ & -                 & $1.5 \times 10^{8}$        & -                                 \\ \hline
\end{tabular}
\caption{\label{tab:1} An overview over the {\sc SHIBONE} simulation suite. In this work we will be analyzing the Full-physics 25 Mpc box suite (bold letters).}
\end{table*}

Accurate theoretical predictions require solving the full nonlinear equations in simulations. Thus, methods to explore the nonlinear regime in \textit{f(R)} gravity are of high theoretical importance. Particularly the nonlinear scales are critical for weak lensing measurement and lend themselves to detailed observational comparisons. 
Quantifying the chameleon effect in detail enables discrimination between different \textit{f(R)} models themselves. To date due to the difficulty solving the coupled scalar field and modified Poisson equations, it has not been straightforward to explore with high resolutions these consequences. In most recent years, several efforts have been done in order to modify existing N-body and hydrodinamical codes to take into account different models of modified gravity. In particular, the $f(R)$ gravity theory  is among the most promising theories (see \citet{Llinares2018} for a review on simulation techniques for modified gravity). 

\subsection{{\sc SHIBONE} simulations}

Baryonic effects in different cosmologies constitute a critical point as theoretical results can be contrasted with observations. A code that can resolve the nonlinear equations of the \textit{f(R)} model but also include baryonics physics is fundamental. The results presented in this work were obtained by analysing the \textsc{SHIBONE} simulation by \citet{Arnold2019_2}. This set of simulations includes a set of full-physics hydrodynamical simulations employing the {\sc Illustris-TNG} model in Hu-Sawicki \textit{f(R)}-gravity \citep{HuSawicki2007}.  

The numerical scheme of this simulation is based on the \textsc{AREPO} \citep{SpringelArepo} code, and employs a new and optimized method to solve the fully nonlinear \textit{f(R)}-gravity equations in the quasi-statics limit, combined with the {\sc Illustris-TNG} galaxy formation model \citep{Pillepich2017, Springel2018,Genel2018,Marinacci2018,Nelson2018}, which incorporates prescriptions for gas-hydrodynamics, star and black hole formation, feedback from supernovae and AGN, magnetic fields, gas heating and cooling processes, as well as galactic winds. The \textsc{shibone} simulations use the same calibration for their baryonic feedback model as the original Illustris-TNG simulations.

The \textsc{SHIBONE} simulations consist of 13 numerical experiments carried out using different cosmologies and at two different resolutions. All simulation initially contain $512^{3}$ dark matter particles (see Table \ref{tab:1}) and the same number of gas cells. For our study, we use the 25 Mpc box, because it has better mass resolution. The simulations start at redshift $z=127$, with a softening length for DM and stars particles of $0.5h^{-1}kpc$. All simulations use Planck 2016 \citep{Planck2016} cosmological parameters $\mathrm{\sigma _{8}= 0.8159}$, $\mathrm{\Omega _{B}= 0.0486}$, $\mathrm{\Omega _{\lambda}= 0.6911}$, $\mathrm{h= 0.6774}$ and $\mathrm{n _{s}= 0.9667}$.

\section{Galaxy selection}
\label{sec:Galaxy_selection}

\subsection{Voids in f(R)}

Voids by definition are underdense regions of the cosmic web. In these regions, due to the low density, potential modifications to gravity should become unscreened and lead to observational differences from GR. 

Such underdense regions provide a powerful tool to investigate the accelerated expansion of the Universe  under a proper environment \citep{Li2012,Cai2015,Paillas2019,Wilson2020,Contarini2021}. The interiors of void regions feature a negative $\delta \rho$ which pushes the $\delta f_{R}$ field to negative values, thereby turning off the screening mechanism and enhancing the modifications of gravity.

Galaxy and CMB surveys have demonstrated how observational data from voids can provide cosmological constrains. Void density profiles, void lensing profiles and redshift spaces distortions are examples of observations that will provide new opportunities to further probe gravity on large scales inside void environments \citep{Li2011,Clampitt2013,Cai2015,Paillas2019}.  

\subsection{Selection of the halo sample}

We aim to investigate the effects of baryonic physics in unscreened and partially screened haloes (\textit{PSH}), where the equivalence principle is no longer valid \citep[see for example the review by][]{Sakstein2020}.

In order to select our halo catalogue, we run a void finders for the whole SHIBONE suite. We applied 3D Spherical void finder \citep[SVF;][]{Paillas2019} which finds spherical voids for a given radius and then, rank the voids in number of increasing neighbours. The outcome is a catalogue of the haloes in the most underdense region of the simulation box. Using the SVF, a complete list of haloes was obtained. For each halo, we get the corresponding number of neighbors in a sphere of 1 Mpc of radius. 

SVF was implemented in order to always have a halo in the void center. The steps to  construct the ranked halo catalogue were:

\begin{itemize}
    \item Gather the total number of halo neighbours for each galaxy. 
    \item Check the local Voronoi cell volume to limit our SVF catalogue to the most underdense regions in the simulated box.
    \item Take the haloes with less neighbours in the most underdense regions, with a cut off mass of  $\mathrm{M_{star}=10^{9}M_{\odot}}$ (See Figure \ref{fig:1A}).
\end{itemize}

\subsection{Reconstructing the gravitational field for the selected haloes}

For the resulting haloes, we mapped the Newtonian potential over the galaxy catalog, according to the \citet{Cabre2012} relations,
\begin{equation}
\begin{matrix}
\mathrm{3\Phi_{\it int}/2c^{2} =  \dfrac{3GM_{200}}{2r_{200}c^{2}}}   \\
\\
\mathrm{3\Phi_{\it ext}/2c^{2} =\sum_{d_{i}< \lambda_{c}+r_{i}}   \dfrac{3GM_{i,200}}{2d_{i}c^{2}}}
\end{matrix}
\label{eq:cabre}
\end{equation}
In these equations, the internal Newtonian potential ($\Phi_{\it int}$) was evaluated using the galaxy mass and the external Newtonian potentials ($\Phi_{\it ext}$) was evaluated using neighbor objects, where $d_{i}$ is the distance to the neighboring galaxy with its corresponding virial mass, $\mathrm{M_{i,200}}$ and virial radius,  $\mathrm{r_{i,200}}$. Finally, $\mathrm{\lambda_{c}}$ is the Compton wavelength given by,
\begin{equation}
\mathrm{\lambda _{C}=32 \sqrt{\left |f_{R0}  \right |/10^{-4}} \mathrm{Mpc}.}
\end{equation}
The \citet{Cabre2012} relations were built to compare the values of the Newtonian fields to the ones of the background scalar field, $f_{R0}$. With this comparison in hand, we can estimate if  galaxies are self-screened ($3\Phi_{\it int}/2c^{2} > \left| f_{R0} \right|$) or unscreened ($3\Phi_{\it int}/2c^{2}< \left|f_{R0} \right| $). The same conditions can be estimated with the external gravity field, $\Phi_{\it ext}$, and the condition to have an environmentally screened regime.

\begin{figure}
\centering
\includegraphics[width=0.8\columnwidth]{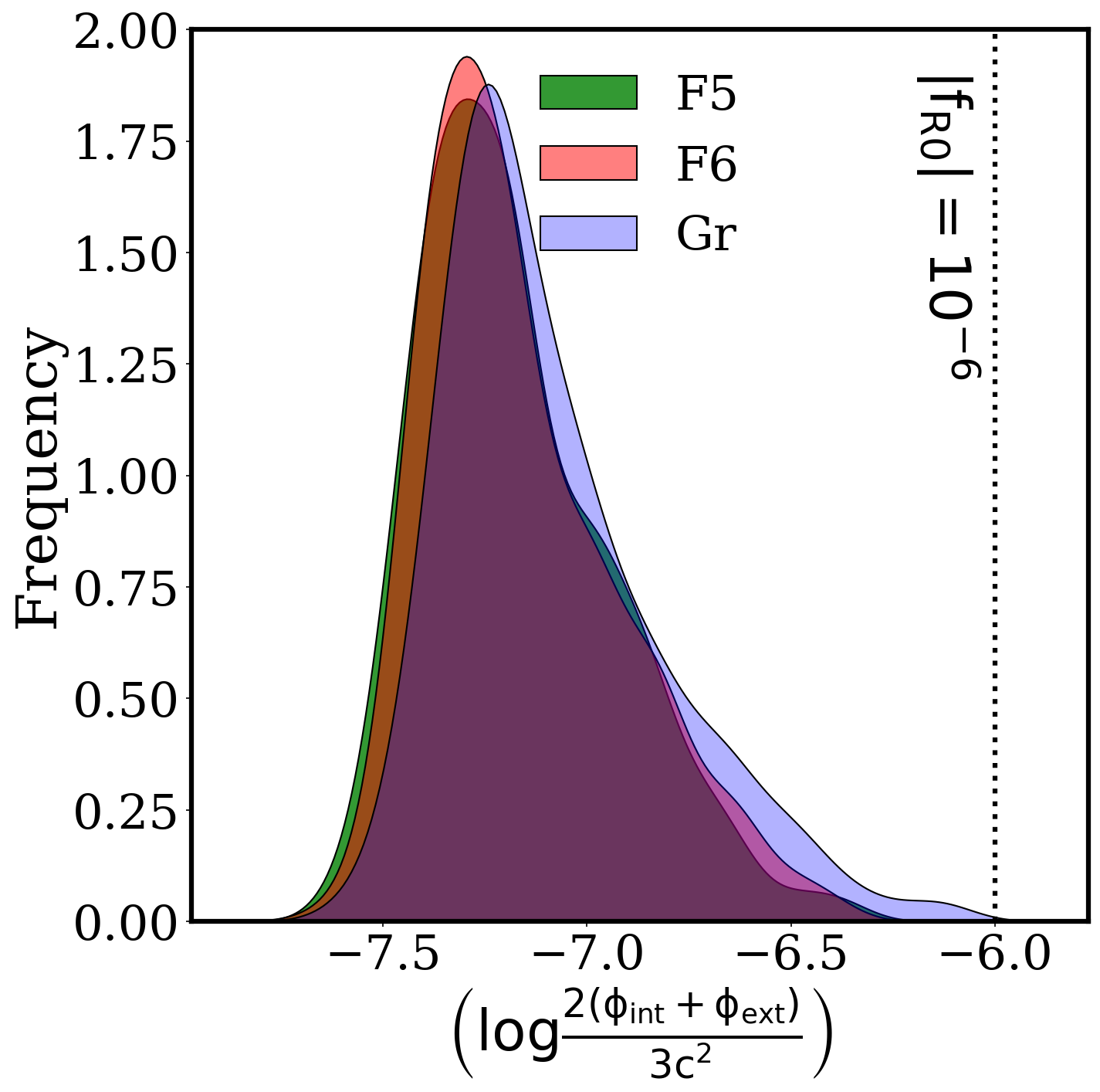}
\caption{The normalised distribution of the total gravitational potential $(\Phi_{\it int}+\Phi_{\it ext})$ in haloes for the three chosen cosmologies (F6, F5 and GR). The vertical dotted line marks the value $\mathrm{|f_{R0}|=10^{-6}}$ for comparison. For nearly all haloes in all our halo catalogues, the \citet{Cabre2012} criterion for unscreened haloes, $\mathrm{3\left(\Phi_{\it int}+\Phi_{\it ext}\right)/2c^{2}< |f_{R0}|}$, is satisfied.}
\label{fig:2}
\end{figure}

The haloes of our selection were not completely self-screened, nor are they environmentally screened. Indeed, for most haloes, both the internal and external contributions to the gravitational potential satisfy the \citet{Cabre2012} conditions for unscreened haloes, $\mathrm{3\Phi_{\it int}/2c^{2}< |f_{R0}|}$ and $\mathrm{3\Phi_{\it ext}/2c^{2}< |f_{R0}|}$, as we can see in Figure \ref{fig:2}.

This general criterion makes the selected haloes suitable to study the baryonic effects in MOG models. According to eqs. (\ref{eq:cabre}), our haloes are at most only partially screened.

\section{Results}
\label{sec:Results}

\subsection{Galaxy properties and Scaling relations}

\subsubsection{Halo density properties}

We divided each halo catalogue in bins of  total mass (gas, stars and dark matter) inside the half-mass radius, $\mathrm{M_{\it tot}[<r_{\it hm}]}$ \footnote{The half-mass radius, $r_{\it hm}$, is defined as the radius that encloses 50 percent of the baryonic mass (gas and star particles)}. Because  $\mathrm{M_{\it tot}}$ also include the contribution from dark matter (dm) particles, these quantities can be used as a characterisation of the concentration of total mass inside galaxies. 

\begin{figure*} 
\centering
\includegraphics[width=0.6\textwidth]{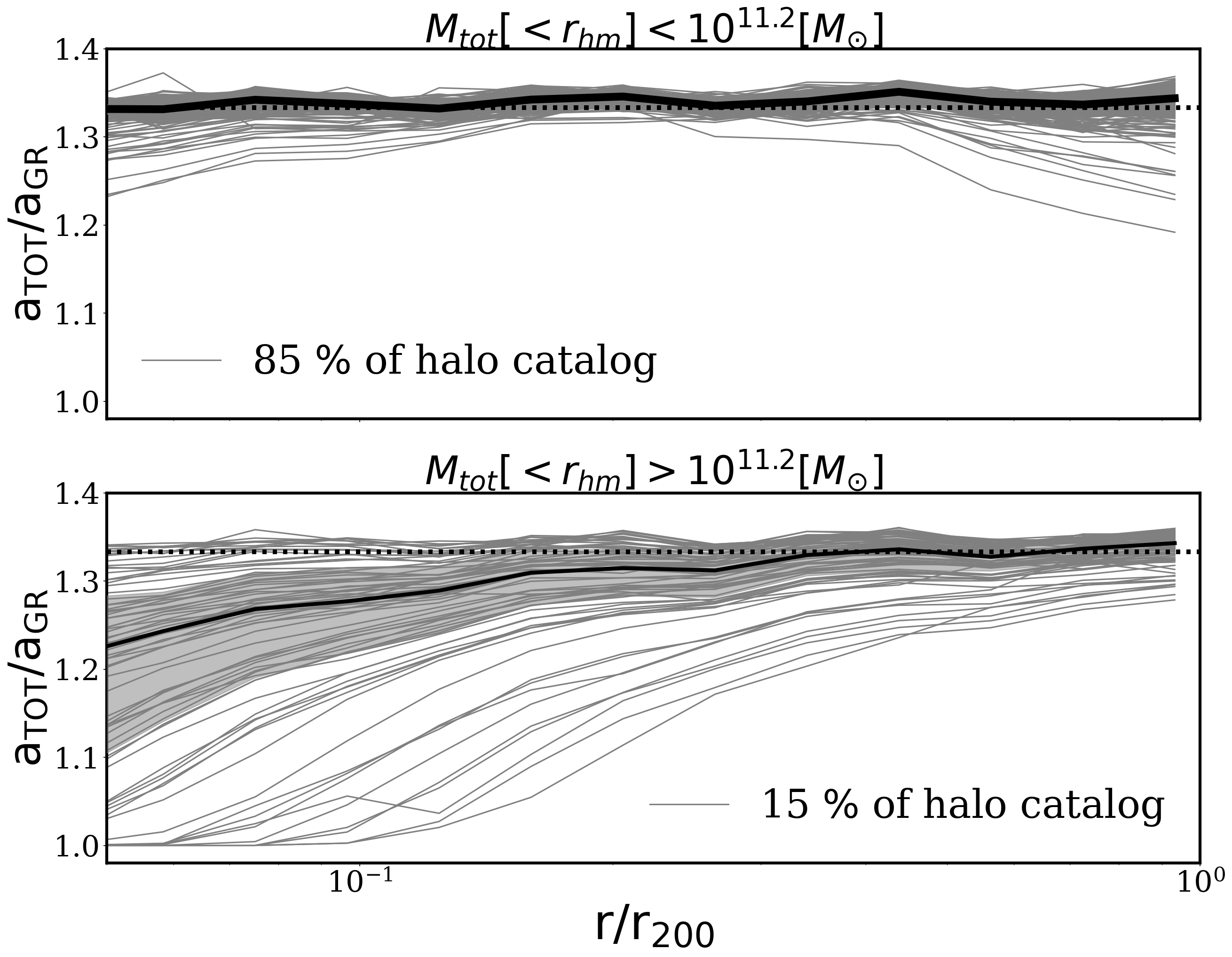}
\caption{The halo acceleration ratio $\mathrm{a_{\it tot}/a_{\it gr}}$ for each selected halo, as a function of the scaled radius, $\mathrm{r/r_{200}}$, for the F6 model in two bins of $\mathrm{M_{\it tot}[<r_{\it hm}]}$ (see the titles of the two subpanels). The ratio for individual haloes is represented as gray lines. The black solid line is the median of the ratio  of all haloes and the shaded areas enclose the $25$th and $75$th  quartiles. For the more massive haloes ($\mathrm{M_{\it tot}[<r_{\it hm}]>10^{11.2}M_{\odot}}$; bottom panel), only their inner parts have screened regions. Haloes with both a screened and an unscreened region are what we define as partially screened haloes (\textit{PSH}). This is approximately $\sim 15 \%$ of the whole halo catalogue. In the case of the F5 simulation, all the haloes are completely unscreened.} 
\label{fig:3}
\end{figure*}

In order to minimize numerical artifacts, we only selected objects resolved with more than $1000$ baryonic particles within the half-mass radius (see the mass resolution in Table \ref{tab:1}). Our goal is to inspect and analyze the ratio between the total and the Newtonian accelerations ($a_{\it tot}/a_{\it gr}$) for each halo, which we show in Figure \ref{fig:3}. This so-called 'acceleration ratio' was studied in the past by \citet{Arnold2016} as a good indicator of the screening radius, beyond which  the effects of the fifth force begin to be relevant, i.e. $\mathrm{a_{\it tot}/a_{\it gr}}$ becomes significantly larger than unity. The acceleration modulus was computed as $\mathrm{(\vec{a}_{x}^2 +\vec{a}_{y}^2 +\vec{a}_{z}^2)^{1/2}}$, for GR accelerations. In the case of the MOG acceleration, the fifth-force contribution was taken into account. 

In the F5 catalogue, the whole halo selection is completely unscreened, as expected of a model with a large background chameleon field. In the case of F6, the total mass inside the half-mass radius seems to be a good indicator of totally unscreened haloes ($\mathrm{M_{\it tot}[<r_{\it hm}] <10^{11.2}M_{\odot}}$) or \textit{PSH} ($\mathrm{M_{\it tot}[<r_{\it hm}]> 10^{11.2}M_{\odot}}$). In the F6 catalogue, \textit{PSH} make up
approximately $\sim 15 \%$ of all haloes.

\subsubsection{Galaxy morphology}

We perform a more quantitative assessment of the demographic of the selected galaxy population, as shown in Figure \ref{fig:4} for the galaxy morphology (see also Figures \ref{fig:1A} and \ref{fig:2A} of the Appendix, for the general scaling relations of the three cosmologies).

To characterise galaxy morphology, we use the disc-to-total stellar mass fraction ratio, ${\rm D/T}$. This criterion was previously implemented by e.g. \citet{Tissera2012,Pedrosa2015} and \citet{Cataldi2020}. This was estimated using the circularity parameter $\epsilon$ of the star particles defined as $\mathrm{\epsilon=J_{z}/J_{z,{\it max}}(E)}$, that is the ratio between the angular momentum $\mathrm{J_{z}}$ and the maximum angular momentum over all particles at a given binding energy $E$, (i.e. $\mathrm{J_{z,{max}}(E)}$). A star on a circular orbit in the disc plane should have $\epsilon \simeq 1$. The disc component is associated with those particles with $\epsilon > 0.5$ and the rest of the particles are associated with the spheroidal component. The ${\rm D/T}$ fraction is the mass fraction in the disc component.

For the central spheroid components (i.e. dispersion-dominated) we define the bulge-to-total fraction as $\mathrm{B/T = 1 - D/T}$.

\begin{figure}  
\centering
\includegraphics[width=0.8\columnwidth]{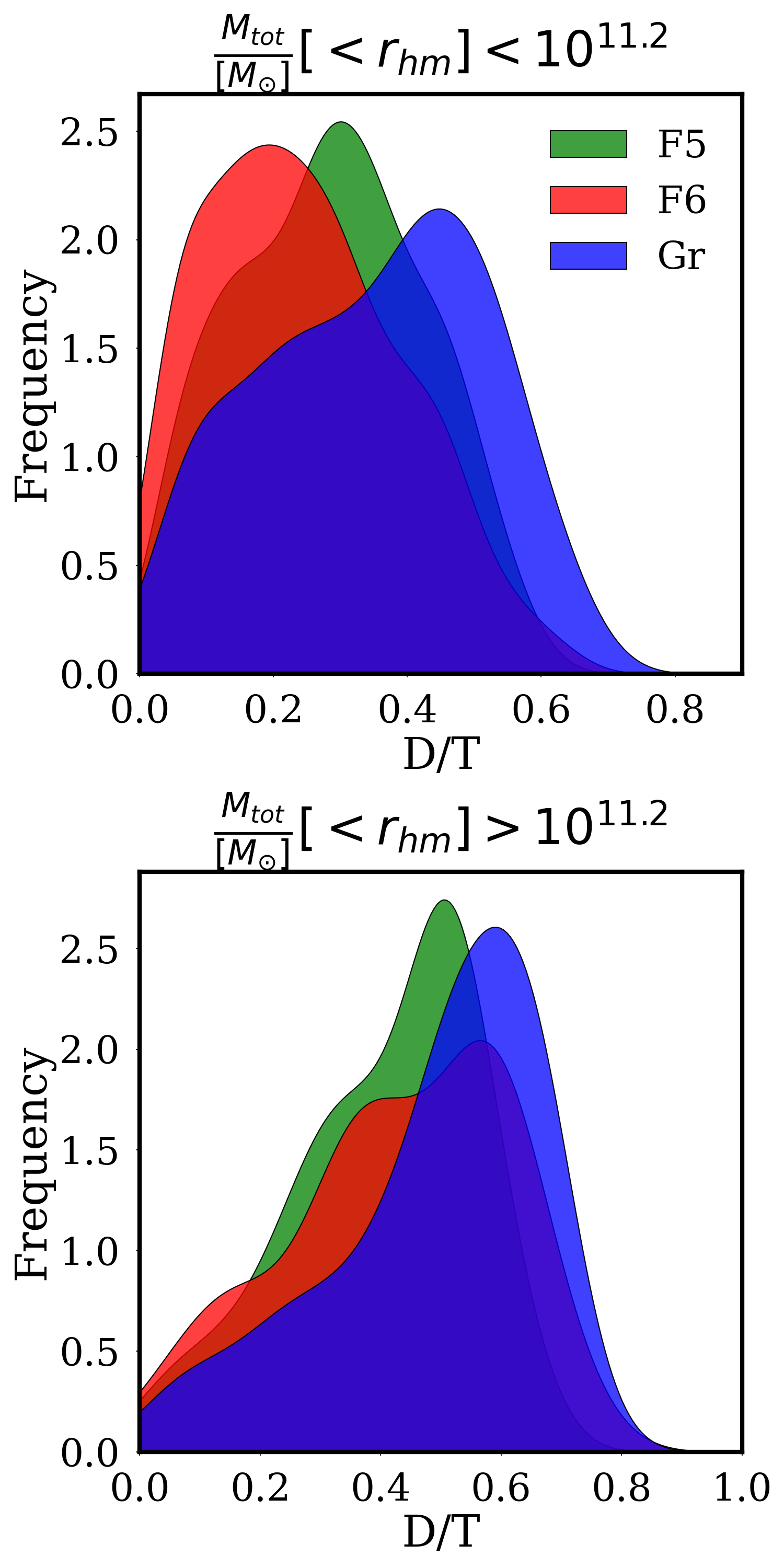}
\caption{(Colour Online) A comparison of galaxy morphology between F6 (red), F5 (green) and GR (blue). Each panel shows a particular mass bin as indicated by the legend. GR contains larger fractions of discs (greater ${\rm D/T}$), and higher frequency of smaller B/T ratios, followed by F5 and  F6, with lower ${\rm D/T}$.}
\label{fig:4}
\end{figure}

The galaxy morphology distribution shows dependence with the mass bins and with cosmologies (see Figure \ref{fig:4}). There is a general trend that the GR catalogue has more well-defined disc-dominated galaxies, followed by F5 and F6, where the elliptical galaxies seem to be the dominant galaxy morphology. 

F6 haloes change significantly across the two mass bins. For galaxies with larger mass within the half baryon mass radius (greater $\mathrm{M_{\it tot}[<r_{\it hm}]}$), galaxies have greater ${\rm D/T}$ fraction. \citet{Arnold2019_2} have reported that {\sc SHIBONE} galaxies can form in $f(R)$-gravity despite the complicated force morphology in the partially screened regime (\textit{PSH}). Even more, there are more well defined disc in F6 compared to GR (and significantly fewer in F5). Following the same trend, we found that MOG haloes increase their ${\rm D/T}$ fraction in the more massive bin, for our smaller halo catalog. 
\subsection{Rotation curves}
\label{sec:observed}

\begin{figure}
\centering
\includegraphics[width=0.51\columnwidth]{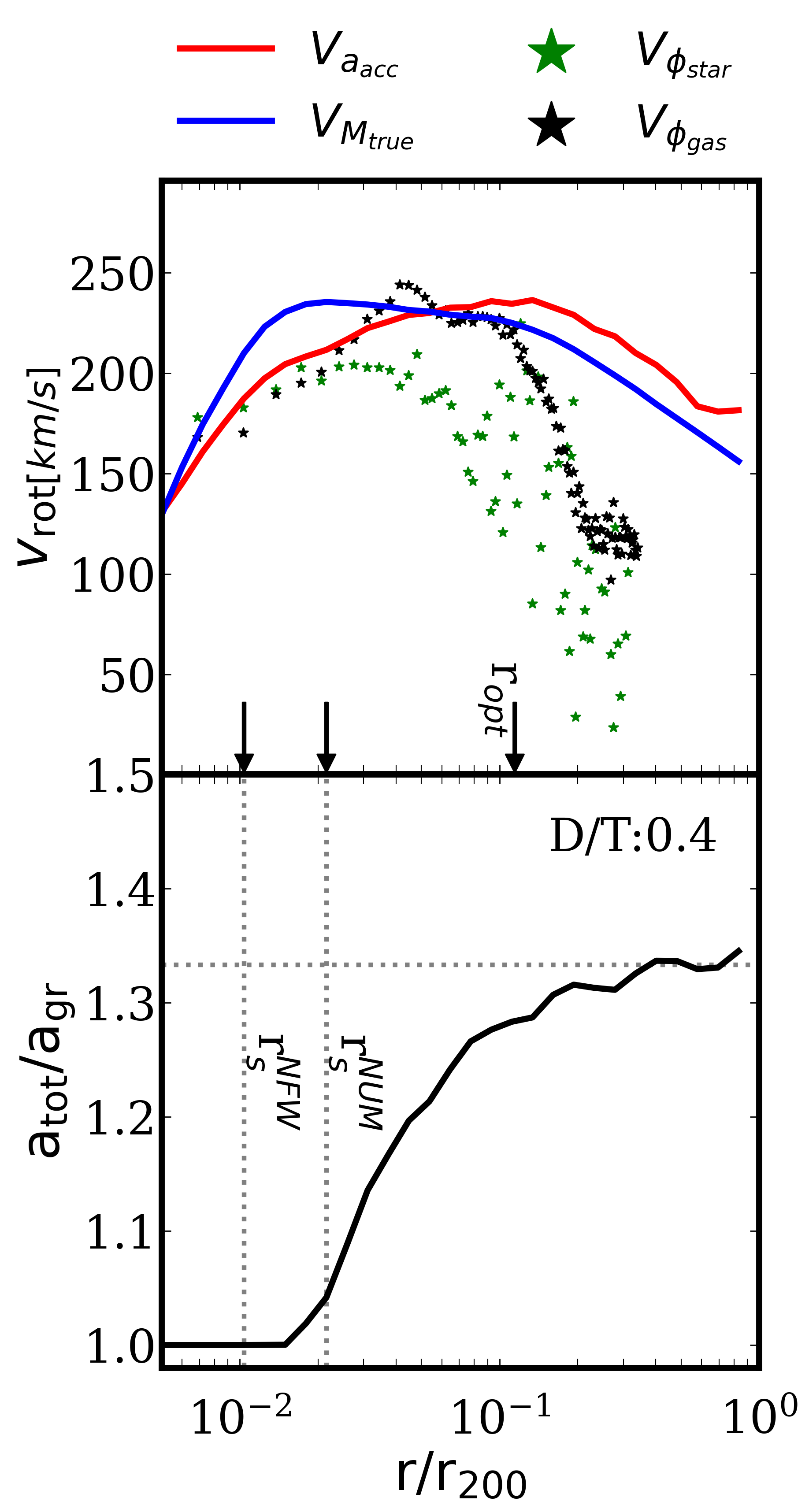}
\includegraphics[width=0.477\columnwidth]{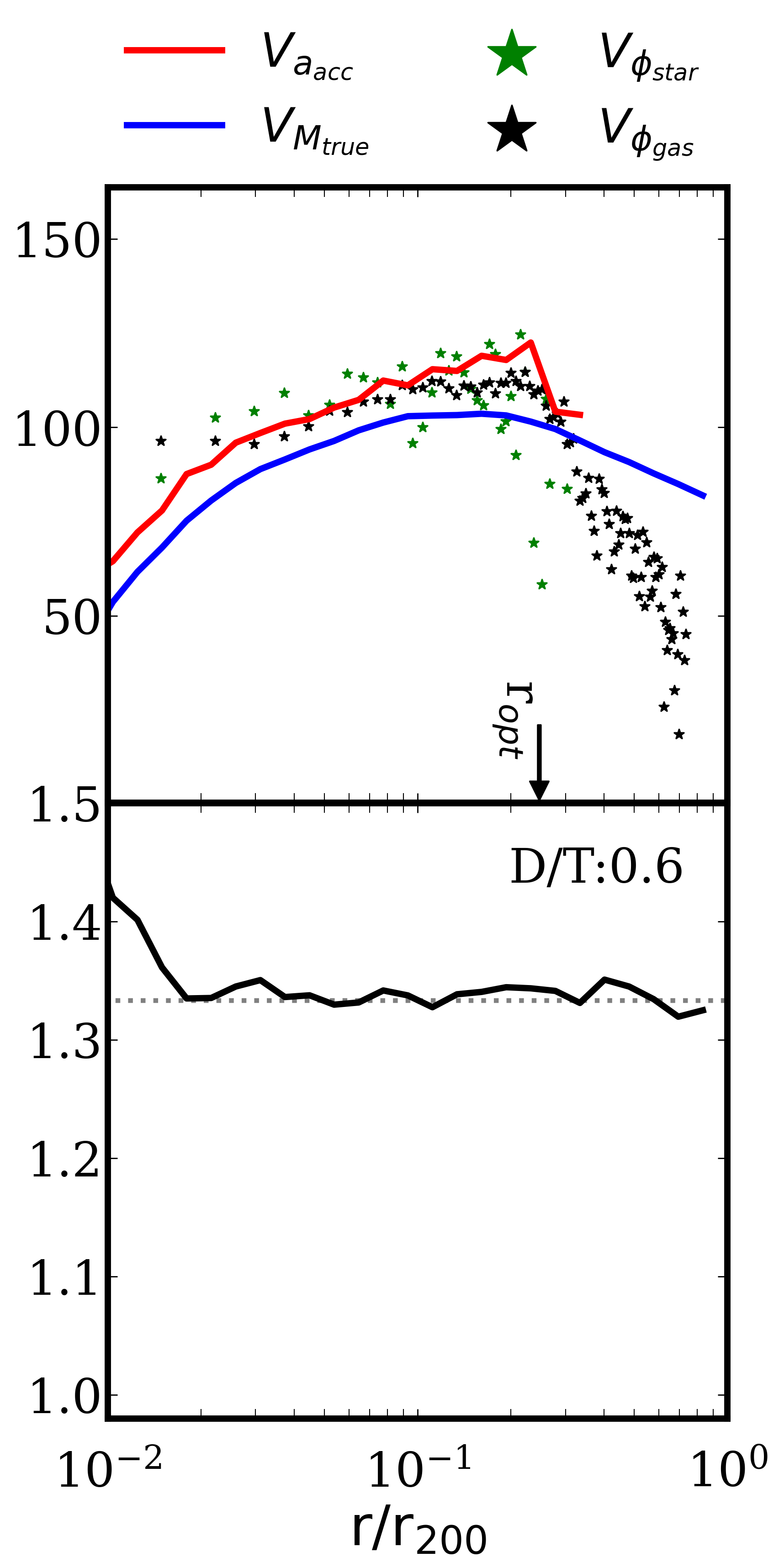}
\caption{(Colour Online) Examples of the observed rotational curves for two F6 haloes: a \textit{PSH} (left panel) and a completely unscreened halo (right panel). $\mathrm{\left \langle V_{\phi,{stars}} \right \rangle}$ and $\mathrm{\left \langle V_{\phi,{gas}} \right \rangle}$ (for stars and gas particles) are shown in green and black star symbols, respectively. We also plotted the $\mathrm{V_{{M_{true}}}}$ (blue line) and $\mathrm{V_{{a_{cc}}}}$ (red line). {\it Bottom panels} the relation  $\mathrm{a_{TOT}/a_{GR}}$ vs $\mathrm{r/r_{200}}$ for the two haloes. The left plot has the screening radius, $\mathrm{r_{s}^{NUM} =1.6[kpc/h]}$, for the acceleration ratios and $\mathrm{r_{ s}^{NFW}=3.3[kpc/h]}$, in vertical dashed lines. Their corresponding values are represented in black arrows in the upper panel. Also, each plot has in black arrow the optical radius, $\mathrm{r_{opt}}$, ($\mathrm{17.4[kpc/h]}$ and $\mathrm{19.8[kpc/h]}$ for right and left panel, respectively), as an estimation where the galaxy disc ends.} 
\label{fig:5}
\end{figure}

To describe the different ways we calculated the baryonic velocities, we illustrate the velocity profiles for two \textit{PSH}s in Figure \ref{fig:5}. We calculate the binned tangential velocity of the stellar disc component. For doing this we choose a system of coordinates perpendicular to the total angular momentum of the galaxy.

Considering only haloes with screening radius less than $\mathrm{r_{opt}}$ \footnote{The optical radius, $\rm r_{\rm opt}$, is defined as the radius that encloses 80 percent of the baryonic mass (gas and stars) of the galaxy.} (the majority of \textit{PSH}s), we calculated the mean tangential velocity of the gas particles, $\mathrm{\left \langle V_{\phi,{\it gas}} \right \rangle}$, in equally spaced radial bins. For systems in rotational equilibrium within such potential wells, we should expect that $\mathrm{\left \langle V_{\phi} \right \rangle \sim  V_{\mathrm{rot}}}$. Finding a departure from this equality could be indicative of a perturbation in the angular momentum by the action of an additional force.

To better visualise the tangential velocities in comparison to $\mathrm{V_{ rot}}$, we analysed two haloes individually for the F6 simulation. In each plot we have indicated the numerical screening radius for the acceleration ratios $\mathrm{r_{s}^{NUM}}$, defined as the radius where $\mathrm{a_{TOT}/a_{GR} = 1.03}$ (see more details about this choice below) and the theoretical screening radius calculated using a NFW density profile, $\mathrm{r_{s}^{NFW}}$ (see eq. \ref{eq:rsNFW}), in vertical grey dashed lines. Both screening radii differ due to the different methods used to calculate them. In the case of \textit{PSH} (left panel) the object is massive enough to affect the relation $\mathrm{a_{TOT}/a_{GR}}$ in the inner radii. The fifth force in this case decrease quickly and the chameleon screening sets in ($\mathrm{a_{TOT} \sim  a_{ GR}}$). 

We analysed the departures between different methods of calculate the rotational velocities, $V_{\mathrm{a_{cc}}}= \sqrt{\boldsymbol{a}_{TOT}\cdot \boldsymbol{r}}$ and $\mathrm{V_{M_{true}} =\sqrt{GM(<r)/r}}$, from the tangential velocity of the disc $\mathrm{\left \langle V_{\phi,{ gas}} \right \rangle}$ and $\mathrm{\left \langle V_{\phi,{ star}} \right \rangle}$(calculated using star particles from the stellar disc). 

The residual velocities $\mathrm{R= \left \langle V_{\phi} \right \rangle - V_{M_{true}}}$ between different methods can be analysed via the Tully-Fisher relation, inspecting the different maximum rotation velocities. In the upcoming years, with MOG simulations with better resolution and for models where gas and stars particles experience different degrees of screening, this kind of plots could be used to check the test proposed by \citet{Vikram2018} in a numerical context.

\subsubsection{Tully-Fisher relations}

The Tully-Fisher relation (TFR) is an empirical law that relates the maximum rotation velocity achieved in the rotation curve of a spiral galaxy and its mass content or luminosity. The TFRs evidence the flattened profiles found in the rotation curves of spiral galaxies (modified from the expected Keplerian falling off curve) by predicting the asymptotic constant rotation velocity of stars far off from the galactic centres in terms of the total mass, or vice versa \citep[e.g, see][]{Acedo2020}.

Modified gravity theories and the TFRs have been connected alongside the first constraint test \citep[e.g.][]{Dutton2009,Trujillo2011, McGaugh2012,Brook2012}, especially in MOND \citep{Milgrom1983} models \citep[e.g.][]{McGaugh2012,Zobnina}. This family of models present a modification of Newton's law of universal gravitation in order to replace dark matter.  Recently, \citet{Amekhyan2021} obtained constraints on \citet{Gurzadyan2019} dark energy model using baryonic TFRs.

The properties of the baryonic TFR (BTFRs) and stellar TFR (STFRs) unavoidably depend on the way the gas and stellar masses are measured. We focus in this work on the stellar relations (STFRs), where the mass can be deduced from the galaxy luminosity with an assumed mass-to-light ratio. The maximum rotation velocity of a galaxy, in a numerical simulation, can be measured or estimated independently with three different methods:

\begin{itemize}
    \item using the integrated total (stars, gas and dark matter) particle mass within radius $r$, as given by $\mathrm{V_{M}=\sqrt{GM(<r)/r}}$;
    \item using the tangential velocity of the gas particles from the gaseous disc, as $\mathrm{V_{\phi,{gas}}}$, where $\mathrm{V_\phi}$ denotes the tangential component of the velocity;
    \item or using the acceleration field in the radial direction, as $\mathrm{V_{\mathrm{a_{cc}}}=\sqrt{\boldsymbol{a}_{TOT}\cdot\boldsymbol{r}}}$, where we have used bold symbols to denote vectors, and $\cdot$ means taking the inner product of two vectors.  
\end{itemize}

The maximum rotation velocity is then taken as the maximum value of the rotation velocity profile, $\mathrm{V_{rot}}$. These three methods are equivalent when the halo is not perturbed by recent mergers, with an intrinsic connection to their halo morphology. More spherical haloes tend to have more similar maximum rotational velocities independent of the method, than more irregular shaped haloes. Mergers  have an important impact on the rotation velocity \citep{Pedrosa2008}. Therefore, for systems in rotational equilibrium within a gravity potential well, we should find the same values for $V_{\mathrm{rot}}$, independently of the calculation method.

\begin{figure*}
\centering
\includegraphics[width=\textwidth]{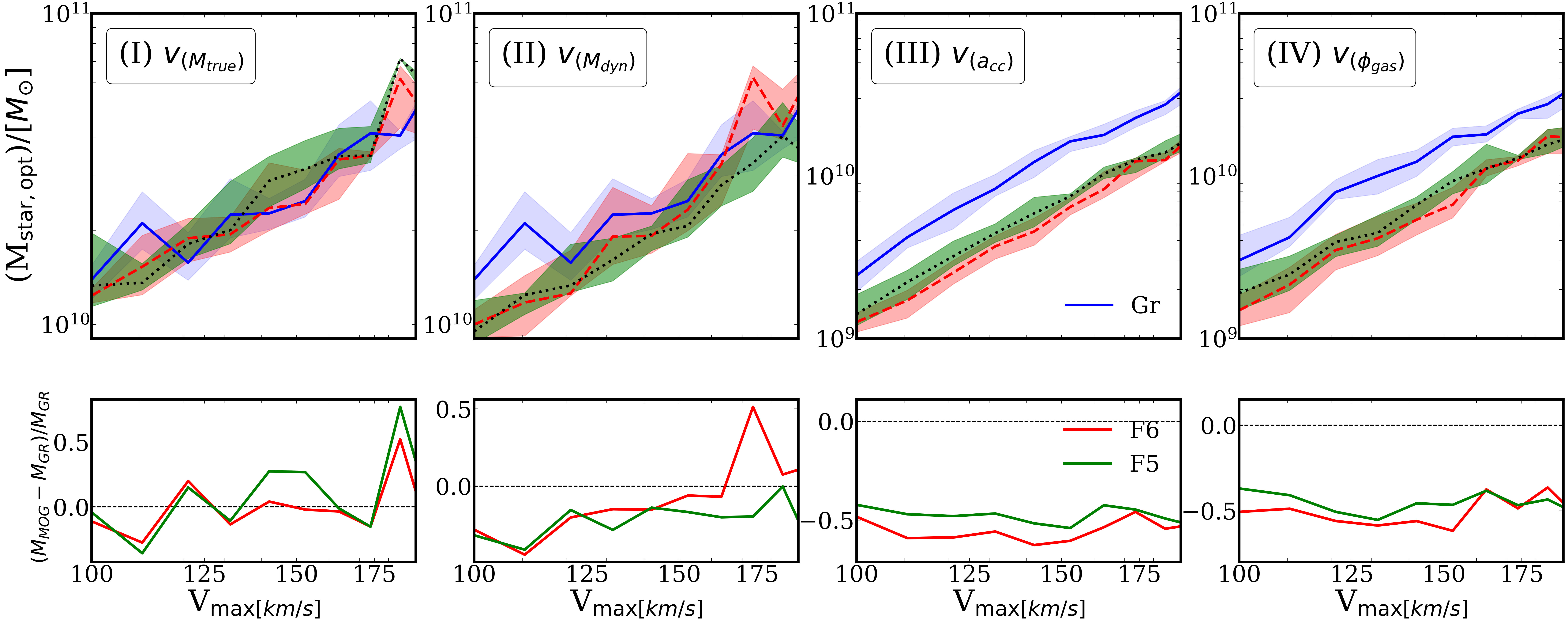}
\caption{(Colour Online) The Stellar Tully Fisher relation (STFR) for the selected haloes. Panel (I) $\mathrm{V_{max}}$ calculated using the integrated mass. Panel (II): $V_{{\it M_{true}}}$,  using the integrated dynamical mass, $\mathrm{V_{{M_{dyn}}}}$ for the MOG models (for GR is equal to  $\mathrm{V_{{M_{true}}}}$). Panel (III) calculated with the total acceleration $\mathrm{V_{{a_{acc}}}}$. Panel (IV) with the tangential velocity of the gas particle, $\mathrm{V_{\phi _{gas}}}$. Bottom panels: the residuals of each MOG cosmology: F6 (red), F5 (green) vs GR (solid blue lines).}  
\label{fig:6}
\end{figure*}

In Figure \ref{fig:6}, we show the STFRs for all three cosmologies and velocity calculation methods. In each panel we show the relative difference in optical mass of each MOG cosmology with respect to GR, i.e. $\mathrm{(M_{MOG}-M_{GR})/M_{GR}}$.

In Panel (I) of Figure \ref{fig:6} the STFR was computed as $\mathrm{V_{{M_{true}}} =\sqrt{GM(<r)/r}}$ and then we took the maximum value. The resulting plot shows no significant differences between the three gravity models. 

Panel (II) is the same calculation using the dynamical mass (discussed later in section \ref{sec:Meff}) as $\mathrm{V_{{M_{dyn}}} =\sqrt{GM_{dyn}(<r)/r}}$. Fixing $\mathrm{M_{star,opt}}$, MOG haloes show larger $\mathrm{V_{max}}$, in comparison with GR. The fifth force does not affect the stellar mass significantly but does increase $\mathrm{M_{dyn}}$ and in consequence, increase $\mathrm{V_{{M_{dyn}}}}$. For more massive galaxies, we recover the behaviour of Panel (I) for F6, as $\mathrm{M_{dyn} \sim M_{true}}$ for the most massive objects in this model. In the case of small haloes in our catalog (that are completely unscreened, see Figure \ref{fig:3}), the difference between GR are significant. This can be clearly seen in the regime of lower $\mathrm{M_{star,opt}}$, where the F6 and F5 small haloes reach the same $\mathrm{V_{max}}$, which is a consequence of the enhancement of the gravitational force.

In Panel (III) we compute the velocity as $\mathrm{V_{{a_{acc}}} =\sqrt{ \boldsymbol{a}_{TOT} \cdot\boldsymbol{r}}}$, where the total acceleration is $\mathrm{a_{GR} + a_{mod}}$. The enhancement of $\mathrm{V_{max}}$ due to the fifth force is in this case direct.

Finally, in Panel (IV) we analyze the velocity as $\mathrm{V_{\phi _{gas}}}$. The differences between models in this case remain significant. The tangential velocities are sensible to recent mergers, overall formation and stability of the stellar disc, in addition to the fifth force. In the case of F6 and F5, different gravity regimes inside the stellar disc, strengths the differences in the  tangential velocity for the same $\mathrm{M_{ star, opt}}$.

When more precise observational determinations of these velocities become available, signs reflecting the effects of MOG could be detected. 

\begin{table*}
\label{tab:2}
\caption{The Stellar Tully-Fisher relation (STFRs) for the three catalogues, with their corresponding Pearson correlation coefficients, $\rho_{pearson} ^{COSMO}$, and their dispersion $1-\sigma$ in the $\mathrm{M_{opt}}$.}
\begin{tabular}{c|c|c|c|c|c|c}
 \textbf{Tully-Fisher (star)} & $\rho_{pearson}^{GR}$ & $\mathrm{\sigma_{\hat{y}}^{GR}[10^{10}M_{\odot}]}$ & $\mathrm{\rho_{pearson} ^{F6}}$ & $\mathrm{\sigma_{\hat{y}}^{F6}[10^{10}M_{\odot}]}$ & $\mathrm{\rho _{pearson} ^{F5}}$ & $\mathrm{\sigma_{\hat{y}}^{F5}[10^{10}M_{\odot}]}$ \\ \hline
$\mathrm{M^{star, opt}(M_{true})}$ vs $\mathrm{V^{max}(M_{true})}$          & $0.87$                 & $2.96 $                                   & $0.91$                 & $1.81 $                                   & $0.90$                  & $1.59$                                    \\
$\mathrm{M^{star,opt}(M_{true})}$ vs $\mathrm{V^{max}(M_{dyn})}$           & $0.87$                 & $2.96$                                    & $0.89$                 & $1.81 $                                   & $0.90$                       & $1.59$                                         \\ 
$\mathrm{M^{star,opt}(M_{true})}$ vs $\mathrm{V^{max}(gas_{tg})}$          & $0.79$                 & $2.97$                                    & $0.86$                 & $1.75$                                    & $0.84$                  & $1.60$                                    \\
$\mathrm{M^{star,opt}(M_{true})}$ vs $\mathrm{V^{max}(a{r})}$              & $0.87$                 & $2.96$                                    & $0.88$                 & $1.81$                                    & $0.90$                  & $1.47$                                    \\ \hline
\end{tabular}
\end{table*}

In Table \ref{tab:2}, we inspect the degree of correlation of $\mathrm{M_{star,opt}}$ vs $\mathrm{V_{max}}$ for each haloes for all analysed cosmologies and velocity methods. For this, we calculated the Pearson coefficient, $\mathrm{\rho_{Pearson}^{COSMO}}$, as a degree of linear correlation between two sets of data. For a totally correlated system this coefficient goes to 1. On the other hand, for uncorrelated sets, the coefficient takes a value equal to 0. In all the cases, the pearson coefficient $\mathrm{\rho_{ Pearson}^{COSMO}}$ reflects strong correlations.

On the other hand, the dispersion $\sigma$ in the optical mass, $\mathrm{M_{ star, opt}}$, for a given $\mathrm{V_{max}}$, varies significantly, giving a possible fingerprint to test MOG effects. 

\subsection{Effective Mass}
\label{sec:Meff}

The dynamical mass of a halo is the mass 'felt' by massive test particles. It can be measured using the relationship between the gravitational potential energy and the kinetic energy of all constituent parts. In the case of simulations, it can be calculated from the density field created by the dark matter particles. More explicitly, the effective density field, $\mathrm{\delta \rho_{ eff}}$, can be defined by casting eq.~(\ref{eq:PoissonMOG}) into the following form \citep{He2015}: 
\begin{equation}
\nabla^{2}\Phi = 4 \pi G \delta \rho_{\it eff},\quad
\delta\rho_{\it eff} \equiv \left(\frac{4}{3}-\frac{\delta R}{24 \pi G \delta \rho}\right)\delta\rho.
\end{equation}
This can be calculated for all the cells in the simulation grid, from which one can calculate a ratio between $\mathrm{\delta\rho_{eff}}$ and $\mathrm{\delta\rho}$. This ratio is multiplied by the mass of all particles residing in that cell in order to calculate an 'effective mass' of these particles. The total effective mass of all particles with the radius of a halo gives the effective mass, $\mathrm{M_{eff}}$. 

$\mathrm{M_{true}}$, the \textit{true} halo mass, is not necessarily the same as $\mathrm{M_{eff}}$. This is defined within the same radius around the same halo centre but using the true mass of particles. \citet{He2015} suggested that it is preferable to use the effective mass for the purpose of analysing the dynamical properties of haloes in $f(R)$ models. $\mathrm{M_{eff}}$ can be used as a proxy for the dynamical mass $\mathrm{M_{dyn}}$; both the effective mass and the dynamical mass vary between $\mathrm{M_{true}}$ and $\mathrm{\frac{4}{3}M_{true}}$: when there is no chameleon suppression of the scalar field, the relation is $\mathrm{M_{ dyn} = \frac{4}{3}M_{true}}$, while when the halo is strongly screened, the dynamical mass reduces to the true value in GR ($\mathrm{M_{dyn}\simeq M_{true}}$). 

\begin{figure}
\centering
\includegraphics[width=\columnwidth]{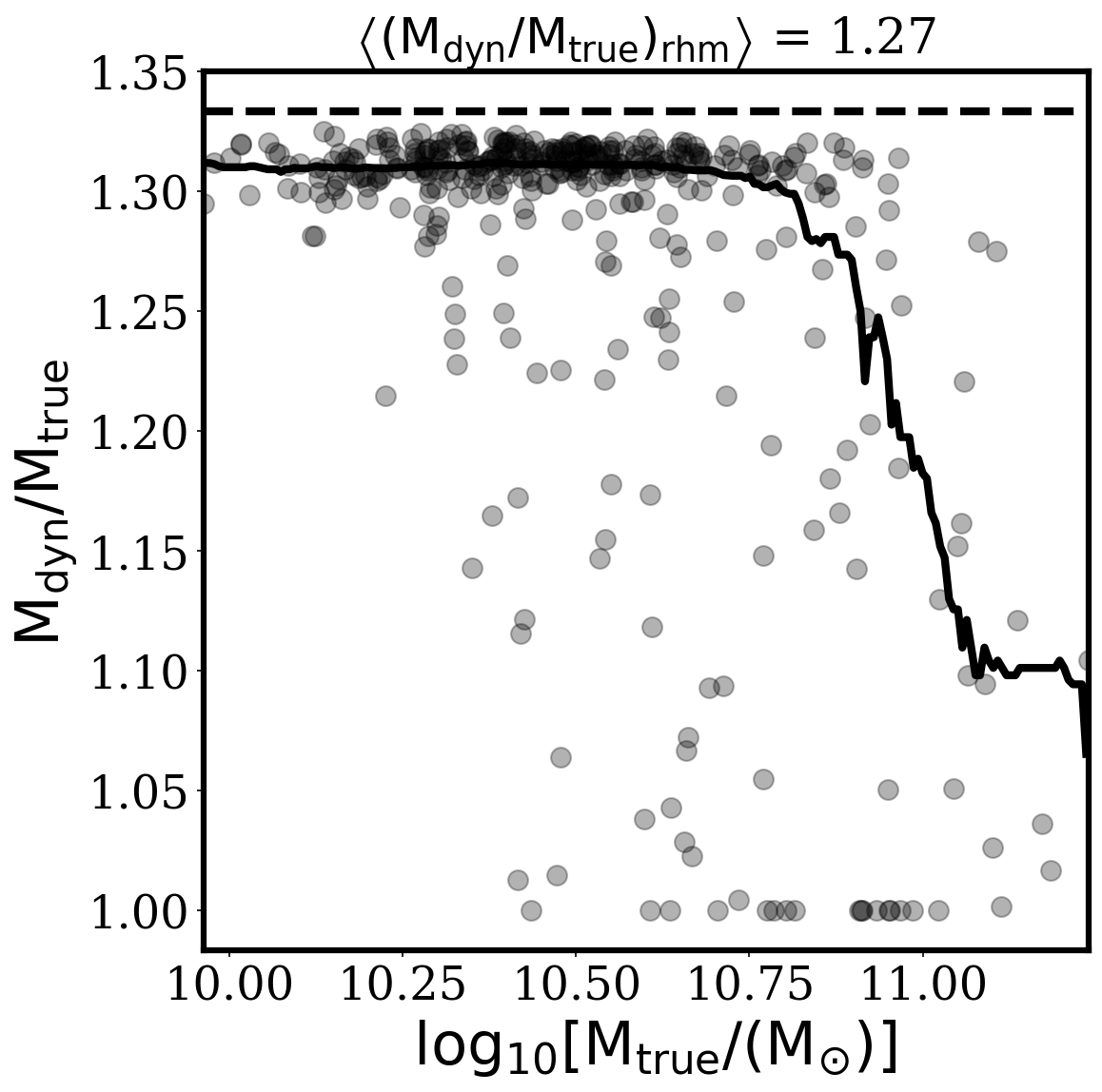}
\caption{The ratio of effective to true masses, $\mathrm{M_{dyn}/M_{true}}$, as a function of the true mass, $\mathrm{M_{true}}$ in the F6 halo catalogue selected based on the total mass inside the half-mass radius, $\mathrm{r_{hm}}$. Each symbol represents an individual halo, and the solid line is the median value. The title of the plot shows the mean value of the mass ratio inside the $\mathrm{r_{hm}}$.}
\label{fig:7}
\end{figure}

Chameleon screening effects come also from the matter that surrounds a halo, commonly known as environmental screening. Due to the conditions imposed for the selection of the sample in this study, we do not have environment effects in the PSH.

In Figure \ref{fig:7} we plot the mass ratio, $\mathrm{M_{dyn}/M_{true}}$, as a function of the true mass, $\mathrm{M_{true}}$, for the F6 haloes. Each point corresponds to an individual halo, where the majority lie along the horizontal line near $4/3$. For the more massive haloes, $\mathrm{M_{ true} \approx M_{dyn}}$, which corresponds to a chameleon screening that is strong enough to suppress the enhancement of the fifth force. 

The legend in Figure \ref{fig:7} indicates the median values of the mass ratio, which shows that up to the half-mass radius, only a small fraction of the F6 haloes is screened, even though we focus on the inner regions of the halo where the screening effect is expected to be stronger. This median value decreases when we analyse the mass within a smaller radius (e.g., $5\%$ of the halo radius: $\mathrm{0.05 \times r_{200}}$), as expected. \citet{Mitchell2018} has proposed, with a good agreement, a `tanh' function 'toy model' with two free parameters to fit the mass ratio.

\subsection{Halo concentration}
\label{sec:cNFW}

In the $\Lambda$CDM model, dark matter haloes are well described by the NFW density profile given by eq.~(\ref{eq:NFW}), which has two free parameters, $\mathrm{\rho_{0}}$ and $\mathrm{r_{_{NFW}}}$. The NFW profile has also been shown to work reasonably well for haloes in $f(R)$ gravity \citep[e.g.,][]{He2015,Arnold2016, Mitchell:2019qke}. Of the two NFW parameters, the scale radius $\mathrm{r_{_{NFW}}}$ can be expressed in terms of the halo concentration, $\mathrm{c_{\Delta} \equiv r_{\Delta}/r_{_{NFW}}}$ and $\mathrm{\rho _{0}}$ can be further fixed using the halo mass. Here, $\Delta$ denotes the mean overdensity within the halo radius, which is commonly used to define the halo radius. For example, $\mathrm{\Delta=200}$ indicates that within the halo radius $\mathrm{r_{200}}$ the mean matter density is $200$ times the critical density of the Universe at the halo redshift.

We study the concentration-mass relation $\mathrm{c_{200}(M_{200})}$ in both screened and unscreened regimes. The halo concentration was originally defined by \citet{Navarro1997} as a parameter of the NFW profile. While there are different methods to calculate it without directly fitting this profile for haloes, the latter is usually a more reliable means of accurately measuring the concentration in a way that is true to its definition. It has been claimed \citep{Mitchell:2019qke} that even in unscreened haloes in $f(R)$ gravity, the concentration can still be measured in the same way, giving a good fit with the NFW profile. 

In $f(R)$ gravity, \citet{Mitchell:2019qke} found that the concentration can become enhanced or reduced due to the effects of the fifth force on the density profile. For haloes which have recently become unscreened, particles experience a greater acceleration due to the stronger gravitational force, while their velocities have not been strongly affected since this process takes time, altering the density profile such that it is raised in the inner regions and lowered in the outer regions. If, on the other hand, a halo has been unscreened for a long time, then the particles speeds have been enhanced by $\simeq1/3$, leading to an increase in kinetic energy that surpasses the deepening of the gravitational potential caused by the fifth force; in such situations the particles tend to move to the outer regions of haloes, decreasing the concentration. There is not yet a general quantitative model for the concentration in $f(R)$ gravity, but \citet{Mitchell:2019qke} provided a fitting formula which works accurately for a wide range of $f(R)$ variants. Similar studies of the effects of the fifth force on the concentration and the density profile can be found for other gravity models \citep[e.g.,][]{Zhao2011b,Lombriser2012,Shi2015,Arnold2016, MitchellDGP2021}.

We cannot make one-to-one comparison of haloes between different cosmologies, as there is no clear correspondence between haloes with different merger histories. In the F6 haloes, we compared the median values of $\mathrm{\Delta c_{200}(M_{200})}$ between F6 and GR for four mass bins $\mathrm{M\left[<r_{ hm}\right]}$. We found that the concentration is greater in F6 than in GR (see Table \ref{tab:cnfw}), in agreement with the findings of \citet{Mitchell:2019qke}.

\begin{table}
\centering
\caption{\label{tab:cnfw} The ratio between the concentrations in F6 and GR ($\mathrm{c_{ F6}/c_{ GR}}$), for four bins of the total mass contained in the half-mass radius, $\mathrm{r_{hm}}$}.
\begin{tabular}{cc}
\hline
\\[-5pt]
$\mathrm{\frac{M}{[M_{\odot}]}[<r_{ hm}]}$       & $\mathrm{c_{ F6}/c_{ GR}}$  \\
\hline
$< 10^{10.3}$   & $1.58$\\

$\left[10^{10.3}, 10^{10.5}\right]$   & $1.28$  \\

$\left[10^{10.5}, 10^{10.7}\right]$   & $1.76$  \\

$>10^{10.7}$   & $1.70$  \\
\hline
\end{tabular}%
%}
\end{table}

\begin{figure}
\centering
\includegraphics[width=\columnwidth]{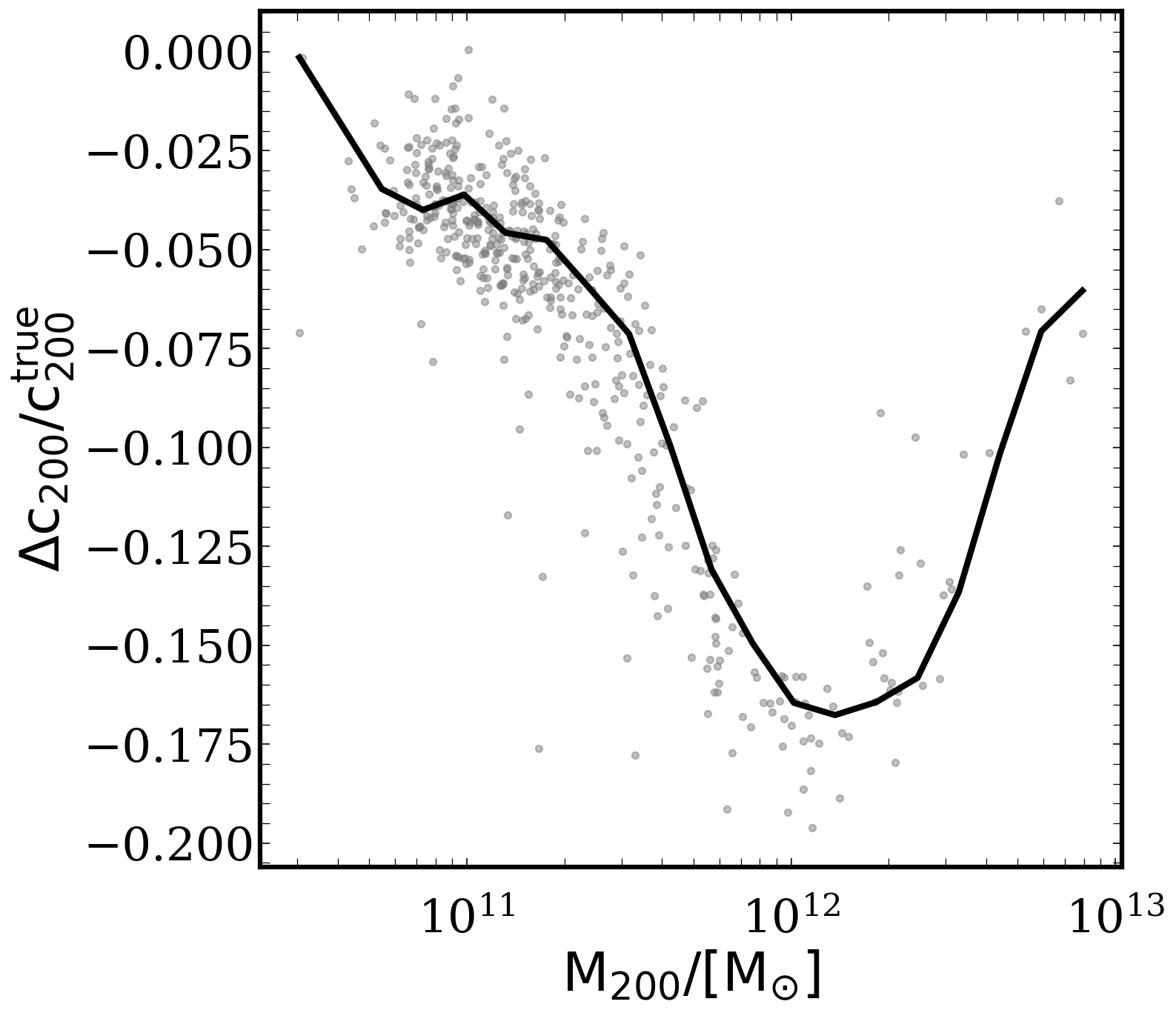}
\caption{The relative difference between the concentration–mass relations based on the effective and true density profiles, at $z = 0$ for the F6 catalogue. The grey points are the individual haloes, and the black solid line shows the moving median relation.}
\label{fig:8}
\end{figure}

We also took the effective density profile, $\mathrm{\rho_{eff}}$, in the F6 haloes and, following the same procedure, we fitted an NFW profile. Figure \ref{fig:8} shows the relative difference between the concentration parameters from the effective and true density profiles, $\mathrm{(c^{dyn}_{F6} - c^{true}_{ F6})/c^{true}_{ F6}}$. For all haloes, we found that when we take into account the additional contribution due to effective mass, the haloes were less concentrated ($\mathrm{c^{dyn}_{F6}< c^{true}_{F6}}$). This effect is as expected. When we consider the dynamical mass, the outer regions, where the fifth force is less screened, receive contribution of an additional term in mass, so that the effective density profile $\mathrm{\rho_{eff}}$ can be significantly higher than the true density profile there. In contrast, in the inner regions $\mathrm{\rho_{ eff}}$ tends to be closer to $\rho$ due to the chameleon screening, and the net effect is a shallower density profile $\mathrm{\rho_{eff}(r)}$ and hence a smaller concentration.

For the more massive haloes, we recovered the true density profile, consistent with our previous findings of Section \ref{sec:Meff}. This should be reflected by tests which aim to measure both the effective and true mass density profile. For example the works by \citet{Terukina2014,Wilcox2015} and \citet{Pizzuti2017} compare the X-ray emitting gas (influence by the fifth force, if it exists) with weak lensing profiles (which recover the true mass) in order to check for a disparity in their contraction.

\subsection{Morphology}

\subsubsection{The screened regions}

The scalar field $f_{R}$ in the innermost regions of a halo can be suppressed by several orders of magnitude with respect to the background field, $\bar{f}_{R}$. This effect is essentially equivalent to switching off the presence of a fifth force. As we go to outer regions, the scalar field grows asymptotically to the value of the background field.

We analysed three different criteria to define a screening radius or the corresponding screening surface. As mentioned in Section \ref{sec:mot}, the screening surface morphology in disc galaxies shows deviations from spherical symmetry, as has been reported by \citet{Naik2018}.

We inspected the face-on and edge-on maps of the scalar field, $f_{R}$, across planes going through the galaxy centres. As an example, see Figure \ref{fig:1}. The contours show the locations of the screening surface for an specific value of the field amplitude ($\mathrm{\left |f_{R}/f_{R0}  \right | = 10^{-2})}$ or total (MOG + GR) vs GR acceleration ratio (i.e. $\mathrm{a_{TOT}/a_{\it GR} = 1.03}$). 

We quantified the deviation of the resulting ellipsoidal screening surface from a spherical morphology, fitting the boundary surface with 2D ellipses for the edge-on and face-on frames independently, according to the equation $\mathrm{{\frac{x^{2}}{ \alpha _{edge}^{2}} + \frac{y^{2}}{ \beta  _{edge}^{2}} = 1}}$ and $\mathrm{{\frac{x^{2}}{ \alpha  _{face}^{2}} + \frac{y^{2}}{ \beta  _{face}^{2}} = 1}}$, respectively. The parameters $\alpha, \beta$ are the major and minor semiaxis of the ellipsoids respectively, where $\it \alpha  \geq \beta$. In addition, we calculated the rotation angle $\mathrm{\phi^{\circ}_{edge}}$ of the ellipses axis with respect to the original stellar disc frame. 

We also fitted the three-dimensional screened region. For this, we use $\mathrm{{\frac{x^{2}}{ \alpha _{ell}^{2}} + \frac{y^{2}}{\beta _{ ell}^{2}} + \frac{z^{2}}{\gamma _{ell}^{2}} = 1}}$ and gather the three parameters $(\alpha _{ell}, \beta _{ell}, \gamma _{ell})$, where $\alpha _{ell}  \geq  \beta _{ell} \geq  \gamma _{ell}$.

To calculate the orientation axis with respect to the stellar disc plane, we considered the Tait–Bryan angles. These angles correspond to the roll, pitch and yaw angles ($\phi,\theta, \psi$) that are defined as the rotation angles around the $\hat{x}$, $ \hat{y}$ and $\hat{z}$ axis, respectively.

\begin{figure}
\centering
\includegraphics[width=\columnwidth]{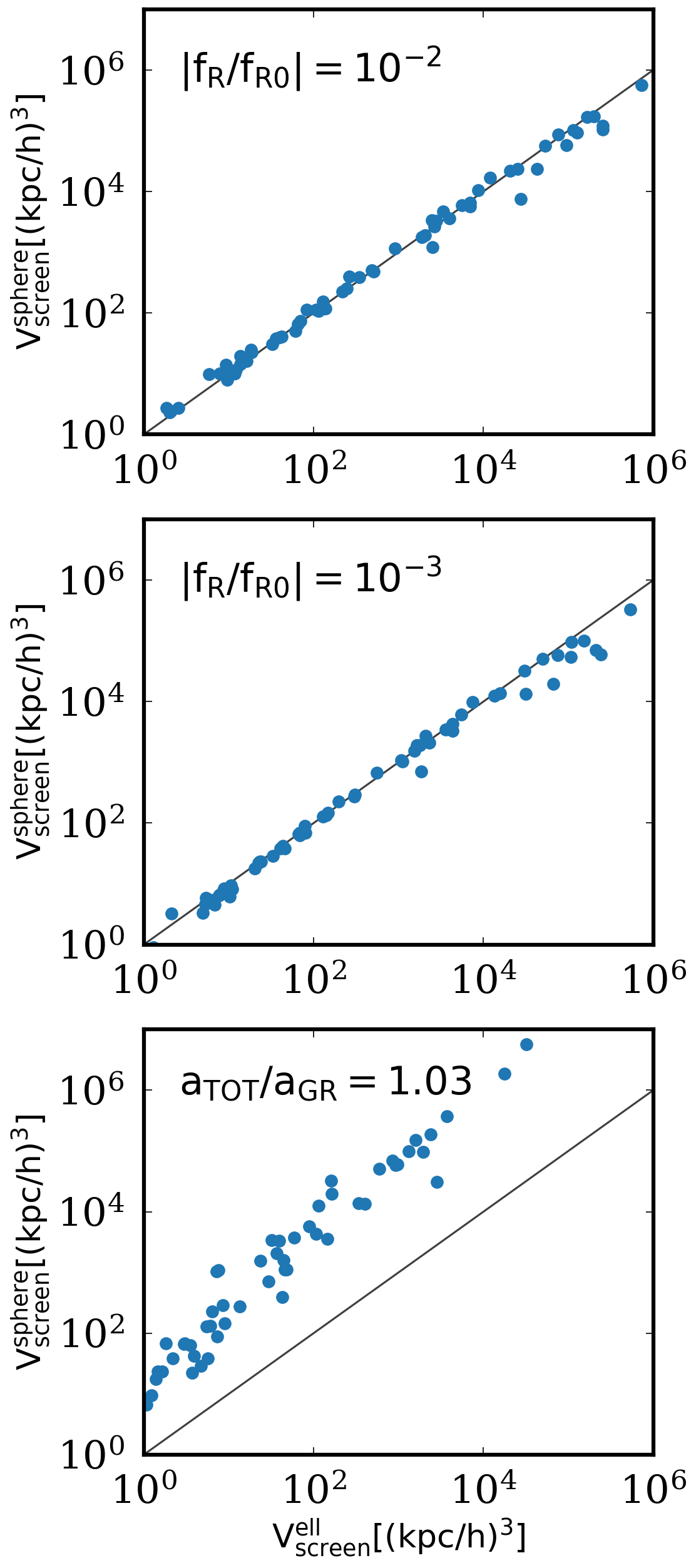}
\caption{Volume fit of the resulting ellipsoids ($\mathrm{\frac{4}{3} \pi  \alpha _{ell} \beta _{ell} \gamma _{ell}}$) vs the equivalent volume ($\mathrm{\frac{4}{3} \pi (r_{s}^{NUM})^{3}}$) of the screening regions, in case we consider a sphere of radius $\mathrm{r_{s}^{NUM}}$. In the latter case, $\mathrm{r_{s}^{NUM}}$ is the radius where the average value shown in the labels of the figure is taken by the field amplitude  $\mathrm{\left|f_{R}/ f_{R0} \right|}$ or the acceleration ratios $\mathrm{a_{TOT}/a_{GR}}$.}
\label{fig:9}
\end{figure}

In Figure \ref{fig:9} we compare the volumes of the ellipsoids versus considering spherical screening regions with a radius equal to $r_{s}^{NUM}$. Labels in each panel shows the criterion to define these screening regions: $\mathrm{\left|f_{R}/ f_{R0} \right| = 10^{-2}}$ (top panel),  $\mathrm{\left|f_{R}/ f_{R0} \right| = 10^{-3}}$ (middle panel) and $\mathrm{a_{TOT}/a_{GR} =1.03}$ (bottom panel). Taking radial bins, we calculate $\mathrm{r_{s}^{NUM}}$ as the radius where field amplitude $\mathrm{\left|f_{R}/ f_{R0} \right|}$ or the acceleration ratios $\mathrm{a_{TOT}/a_{\it GR}}$ takes the average value of the one shown in the labels of the figure. 

For $\mathrm{\left|f_{R}/ f_{R0} \right| = 10^{-2}}$ and $\mathrm{\left|  f_{R}/ f_{R0} \right| = 10^{-3}}$, the resulting volumes (ellipsoids vs spheres) share similar values, deviating from the equality only for the more massive haloes which corresponded to the larger screening volumes. Interestingly, the bigger differences were found when $\mathrm{a_{TOT}/a_{GR} =1.03}$ was the criterion adopted. In order to conserve the screening volumes, from this point on, we adopt $\mathrm{\left|  f_{R}/ f_{R0} \right| = 10^{-2}}$ as the main criterion to define the edges of the screening volumes. 

Even though the volumes were similar whether we parametrize with only one parameter $\mathrm{r_{s}^{NUM}}$ (sphere) or with three parameter $\alpha _{ell},  \beta _{ell},\gamma _{ell}$ (ellipsoids), in the latter case the three ellipsoidal parameters are quite different from each other (see Table \ref{tab:tab_parameter} and Figure \ref{fig:3}).

Stellar disc mass distribution of the galaxy reshapes the screening volume morphology. To quantify this interdependence, we studied how the ellipsoids (3D) and ellipses (2D) parameters behave in terms of $\textrm{D/T}$ bins.
\begin{table}
\centering
\caption{\label{tab:tab_parameter} The mean values of the shape parameters ratios and the angle rotation from the stellar disc frame vs D/T bins.}
\resizebox{\columnwidth}{!}{%
\begin{tabular}{ccccc}
\hline
\\[-5pt]
 & $\left< \mathrm{\beta _{ face}/\alpha _{face}} \right>$  & $\left< \mathrm{\beta _{edge}/\alpha _{edge}} \right>$ & $\left< \phi^{\circ}_{edge} \right>$ & \\
\hline
$\mathrm{0.0<D/T<0.3}$   & $0.89$ & $0.68$ &  $-0.7^{\circ}$ &\\

$\mathrm{0.3<D/T<0.6}$   & $0.87$  & $0.68$& $-2.4^{\circ}$ &\\

$\mathrm{0.6<D/T<1.0}$   & $0.84$ & $0.64$ & $-6.0^{\circ}$ &\\

\hline

\\[-5pt]

 & $\left< \mathrm{\gamma _{ell}/\alpha_{ell}} \right>$  & $\left< \mathrm{\gamma _{ell}/\beta _{ell}} \right>$ & $\left< \phi \right>_{roll}$ &  $\left< \theta \right> _{pitch}$\\
\hline
$\mathrm{0.0<D/T<0.3}$   & $0.59 $   &  $0.70$ & $7.8^{\circ}$ & $-2.3^{\circ}$ \\

$\mathrm{0.3<D/T<0.6}$   & $0.56$  & $0.69$ & $-15.7^{\circ}$ & $-1.2^{\circ}$\\

$\mathrm{0.6<D/T<1.0}$   & $0.55$  & $0.66$ & $6.7^{\circ}$ & $10.0^{\circ}$ \\

\hline
\end{tabular}%
}
\end{table}
The plane containing the major axis of the ellipsoids (i.e $\mathrm{\alpha _{ell}, \beta _{ell}}$) deviate very little from the stellar disc frame (i.e. small $\mathrm{\left< \phi \right>_{roll}}$ and $\mathrm{\left< \theta \right> _{pitch}}$). The fitted regions are aligned to the stellar disc (see Table \ref{tab:tab_parameter}). 

The deviation from a sphere should be reflected in the ratio between axis along the stellar disc. We plotted the relation between ellipse parameters  in Figure \ref{fig:3A} of the Appendix section and are listed in Table \ref{tab:tab_parameter}. In the face-on and edge-on frame, the shape parameter ratio ($\mathrm{\beta _{face}/\alpha_{face}}$) anticorrelates with $\mathrm{D/T}$, i.e. as the stellar disc is more well-defined. The same behavior was found for the  ratio $\mathrm{\gamma _{ell}/\alpha _{ ell}}$ which decreases with $\mathrm{D/T}$. The trend, albeit weak, qualitatively follows what was seen in past studies \citep[e.g.][]{Arnold2016, Naik2018}, namely, that the screening surface loses spherical symmetry and becomes more elliptical, as the stellar disc becomes more well defined.  

\subsubsection{The MOG halo morphology}

We also study the halo shapes and galaxy morphologies. We describe their shapes using the semi-axes of the triaxial ellipsoids, $\mathrm{ a \geq b \geq c}$, where a, b and c are the major, intermediate and minor axes respectively of the reduced moment of inertia tensor, $\mathrm{ S_{ij} = \sum_{k}\frac{r_{k,i}r_{k,j}}{r_{k}^{2}}}$  \citep[e.g.][]{Bailin, Zemp}, where the sub-index represents each mass unit.

To obtain the ratios $\mathrm{q\equiv b/a}$ and $\mathrm{s\equiv c/a}$,  we diagonalized $\mathrm{S_{ij}}$ to compute the eigenvectors and eigenvalues, as described in \citet{Tissera1998}. An iterative method is used, starting with particles selected in a spherical shell \citep[i.e. $ \mathrm{ q=s=1}$][]{Dubinski, Curir1993}. Traditionally the $s$ shape parameter has been used as a measure of halo sphericity \citep[e.g.][]{Allgood,Vera-Ciro, Chua}.

We adopt the triaxiality parameter, defined as $\mathrm{T \equiv (1-q^{2})/(1-s^{2})}$, which quantifies the degree of prolatness or oblatness: $\mathrm{T=1}$ describes a completely prolate halo ($\mathrm{a > b\approx c}$) while $\mathrm{T = 0}$ describes a completely oblate halo ($\mathrm{ a\approx b > c}$). Haloes with $\mathrm{T > 0.67}$ are considered prolate and haloes with $\mathrm{T < 0.33}$ oblates, while those with $\mathrm{0.33 <T <0.67}$ are considered triaxials \citep{Allgood,Artale2018}. DM haloes morphologies have been found to be significantly non-spherical in the N-body simulations \citep[e.g.][]{Jing2002, Allgood,Maccio2008,Vera-Ciro,Despali2014}, and found to be well characterized as triaxial ellipsoids. 

\begin{figure*}
\centering
\includegraphics[width=1.0\textwidth]{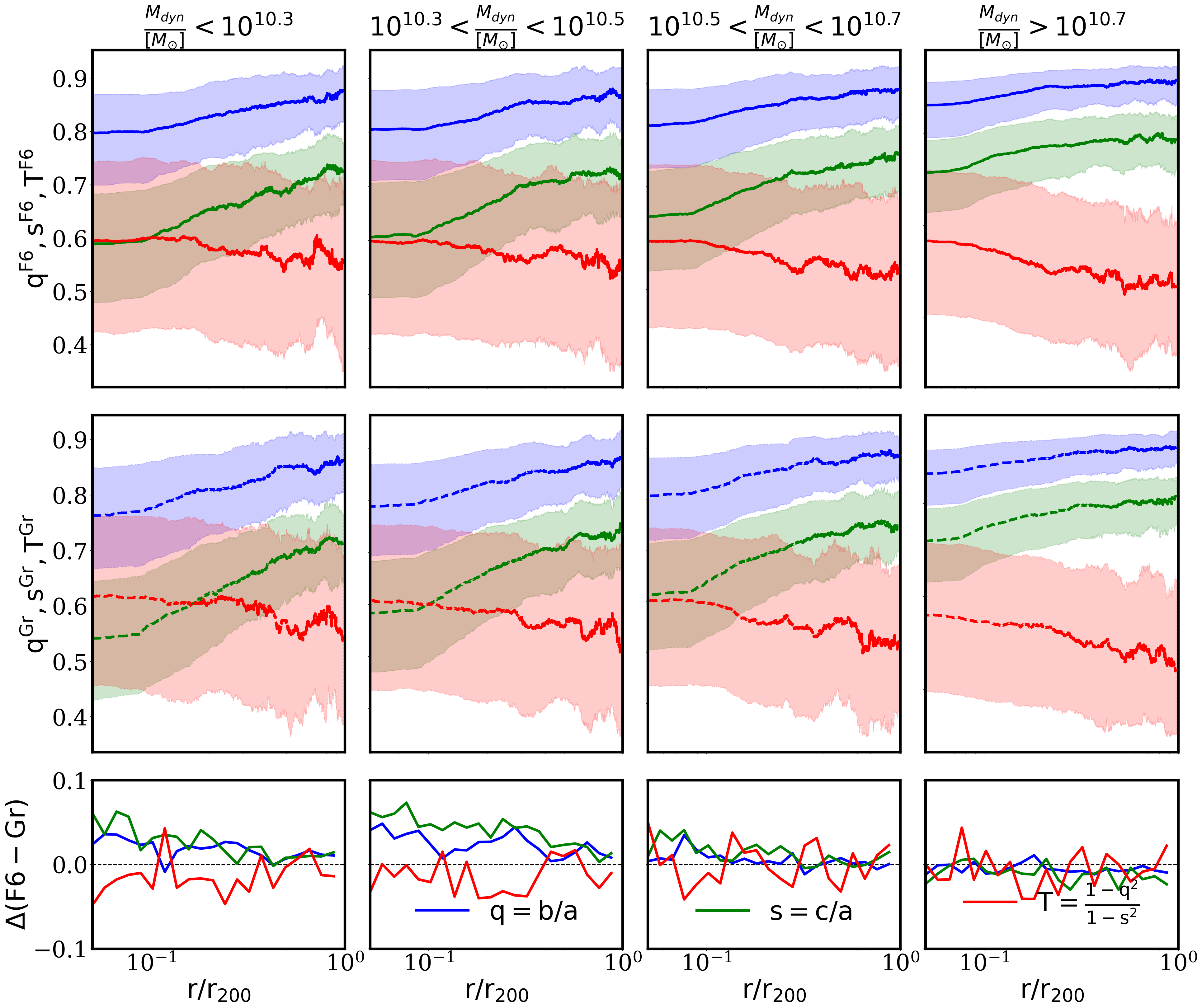}
\caption{(Colour Online) Shape parameters q (blue), s (green) and T (red) median values versus $\mathrm{r/R_{200}}$ for F6 (top panels) and GR (middle panels). The shaded areas enclose the 25th and 75th quartiles. Haloes have been divided in four subsamples according to the total effective mass, shown in the four different columns. The bottom panel of each column shows the relative change between the shape parameters of the F6 and GR runs.}
\label{fig:10}
\end{figure*}

Figure \ref{fig:10} shows the median shape parameters for F6 (top panels) and GR (middle panels), divided in mass bins. In each mass bin, we include the relative change between cosmologies  (bottom panels). F6 haloes are more prolate and less triaxial than their GR counterparts for the less massive bins. Even though this trend is very weak, we note that the morphology of DM haloes is poorly constrained and until now a study of the cosmology dependence is still necessary, which makes the results here useful. The triaxial shapes of haloes have been found, in theoretical and observational studies, to exhibit weak trends with environments, with haloes in underdense environments and of higher mass being more prolate, $\mathrm{T > 0.67}$  \citep[e.g.][]{Maccio2007,vanUitert2017, Lee2017,Gouin2021,Hellwing2021,Menker2022}. F6 haloes in void environments reinforce the trend observed in GR but with the differences between cosmologies becoming smaller when  F6 haloes start to become screened (more massive bins) in the inner regions. 

\FloatBarrier
\FloatBarrier

\section{Conclusions}

In this work we have investigated the impact of alternative gravities (MOG) on  dark matter haloes and their baryons using a statistical approach.  We found  general trends by comparing  simulation boxes with same initial conditions but with different gravities.  Our main results can be summarised as follows.

\begin{itemize}
    \item  The stellar Tully-Fisher Relations show differences between cosmologies when we consider $\mathrm{V_{max}}$ calculated using the acceleration fields and the disc gas tangential velocities. For a constant stellar mass, F6 and F5 tend to have greater maximum velocities, product of the enhancement of the gravity force. More precise observational determinations of these velocities are necessary to be able to distinguish between astrophysical and cosmological effects.
    \item In MOG cosmologies, haloes are more concentrated than in GR. If we look at the profiles of the dynamical halo mass, the concentration decreases in comparison to the GR density profile, as more mass is added in the outer regions, where the halo is unscreened.
    \item The stellar disc interacts with the overall potential well in the central regions, modifying the morphology of the screening regions. Galaxies with greater $\mathrm{D/T}$ fractions deviate more from the spherical shape (even though the spherical volume is conserved) in the sense that stellar disc contracts or elongates the screening shape axes. We also find that the resulting  major axis of the ellipsoid is aligned with the stellar disc.  
    \item Small F6 haloes are less triaxial and more prolate than their GR counterparts. The difference between shape parameters becomes smaller when the F6 haloes start to become screened in their inner region, which becomes more common as the mass of the haloes increases. 
\end{itemize}
These results indicate that careful measurements of lensing masses and shapes, combined with measurements of circular velocities for individual objects to avoid differences in expected concentration vs. mass relations, could be combined in future studies in order to further test and search for modified gravity cosmologies.

\label{sec:conclusions}

\section*{Acknowledgements}

This project has received funding from the European Union’s Horizon 2020 Research and Innovation Programme under the Marie Skłodowska-Curie grant agreement No 734374 and the GALNET Network (ANID, Chile). Also this project was supported through PIP CONICET 11220170100638CO; SP acknowledges partial support by  the Ministerio de Ciencia, Innovación y Universidades (MICIU/FEDER) under research grant PGC2018-094975-C21. SL is supported by PIP 11220200100729CO and grant 20020170100129BA UBACYT. CA and BL are supported by the European Research Council via grant ERC-StG-716532-PUNCA. BL is additionally supported by STFC Consolidated Grants ST/T000244/1 and ST/P000541/1. This work used the DiRAC Data Centric system at Durham University, operated by the Institute for Computational Cosmology on behalf of the STFC DiRAC HPC Facility (www.dirac.ac.uk). This equipment was funded by BIS National E-infrastructure capital grant ST/K00042X/1, STFC capital grants ST/H008519/1 and ST/K00087X/1, STFC DiRAC Operations grant ST/K003267/1 and Durham University. DiRAC is part of the National E-Infrastructure.
%%%%%%%%%%%%%%%%%%%%%%%%%%%%%%%%%%%%%%%%%%%%%%%%%%
\section*{Data Availability}

The data underlying this article will be shared on reasonable request to the corresponding authors.

%%%%%%%%%%%%%%%%%%%% REFERENCES %%%%%%%%%%%%%%%%%%
% The best way to enter references is to use BibTeX:
\bibliographystyle{mnras}
% here we change the meaning of \VAN to use the prefix for the bibliography
\DeclareRobustCommand{\VAN}[3]{#3}
\bibliography{Cataldietal} % if your bibtex file is called example.bib

%%%%%%%%%%%%%%%%% APPENDICES %%%%%%%%%%%%%%%%%%%%%
\clearpage
\appendix
\label{sec:appendix}

\setcounter{table}{0}
\renewcommand{\thetable}{A\arabic{table}}
\setcounter{figure}{0}
\renewcommand{\thefigure}{A\arabic{figure}}

\section{Construction of the catalogue using Voronoi tessellation.}
\label{sec:Voro}

\textsc{Voro++} is an useful numerical code to compute a three-dimensional Voronoi tessellation in a coordinate space. We constructed a complementary catalogue for the three simulations (F6, F5 and GR) and selected the haloes with an environment corresponding to  the lowest Voronoi densities (a degree of how isolated are in the environment). We plotted the results in a 1D histogram that shows where the chosen haloes reside in terms of Voronoi cell volume. We compared the resulting \textsc{Voro++} catalogue with the haloes selected with the SVF method.

\FloatBarrier
\begin{table}
\centering
\label{tab:3}
\begin{tabular}{llll}
\hline
Cosmology  & GR overlap   & F6 overlap & F5  overlap \\ \hline
Taking 100 haloes  & 23 $(23 \%)$  &  13 $(13 \%)$  & 17 $(17 \%)$  \\
Taking 200 haloes  & 98 $(49 \%)$  &  75 $(37.5 \%)$ & 76  $(38 \%)$ \\
Taking 300 haloes  & 215 $(72 \%)$ &  171 $(57 \%)$& 188  $(63 \%)$ \\
Taking 400 haloes  & 386 $(97 \%)$ &  318 $(80 \%)$ & 332 $(83 \%)$  \\\hline
\end{tabular}
\caption{An overview over the number of overlapping haloes between the two methods to construct halo catalogues used in this work.}
\end{table}

As we can see in the Table \ref{tab:3}, the overlap between the two catalogues was considerable. We kept the SVF method for constructing the halo catalogues, as  both methods select nearly the same haloes (see Table \ref{tab:3}).

\section{Properties and Scaling relations}

\begin{figure*}
\centering
\includegraphics[width=0.8\textwidth]{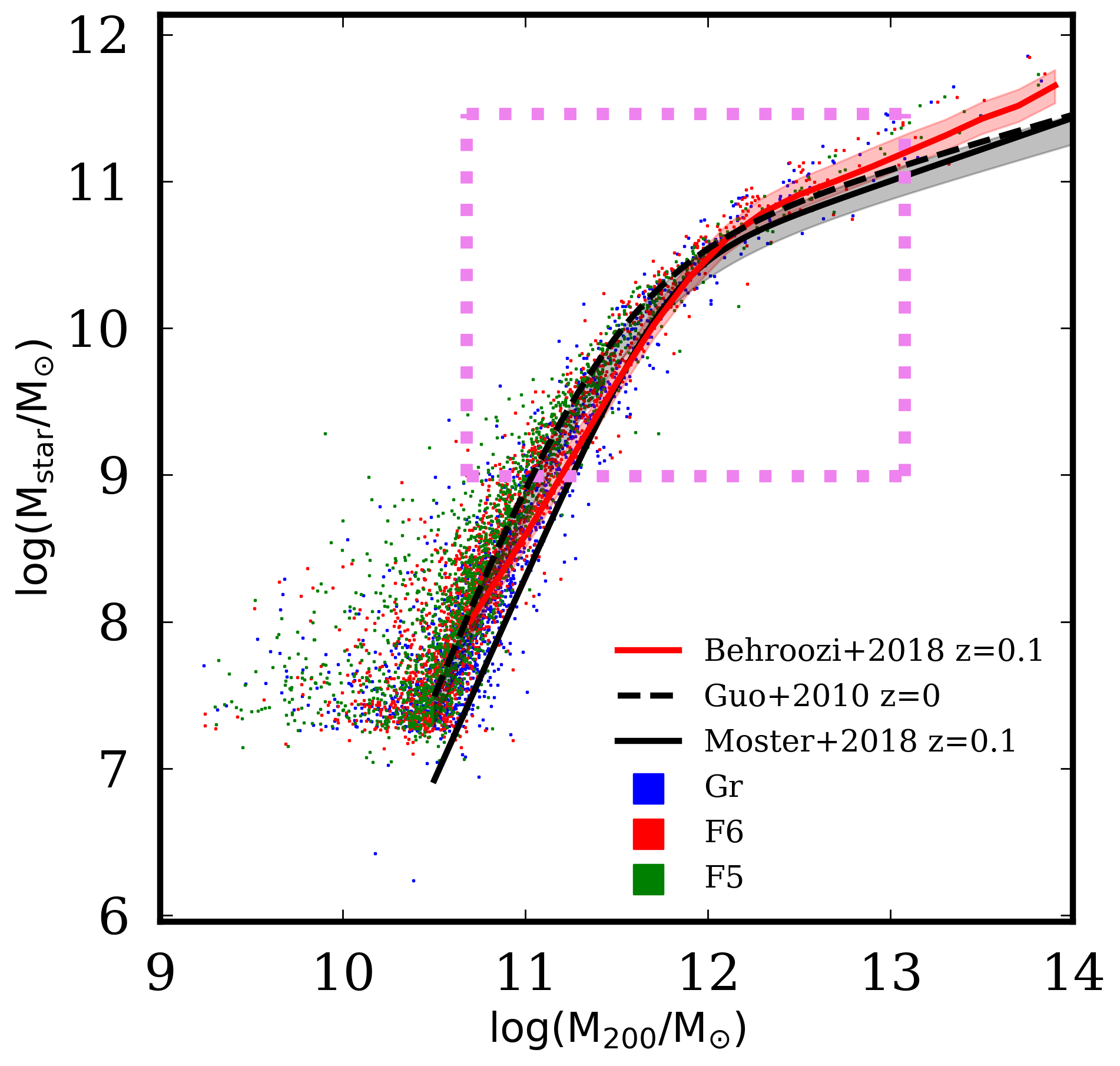}
\caption{The SMHM relation, defined as the ratio of galaxy stellar mass to halo mass for the selected haloes, for the GR simulation (blue), F6 simulation (red) and F5 run (green). The red dashed square shows the stellar and halo mass cuts used to select our halo catalogue. We also show the \citet{Moster2018}, \citet{Guo2010} and \citet{Behroozi2013} models.} 
\label{fig:1A}
\end{figure*}

In Figure \ref{fig:1A} we show the SMHM relation, including the \citet{Moster2018}, \citet{Guo2010} and \citet{Behroozi2013} models.

For the three simulations the relation has similar values as were already present in the original {\sc Illustris-TNG} simulation (see Figure 4 of \citet{Pillepich2017}) and also reported by \citet{Arnold2019}.  

\begin{figure*}
\centering
\includegraphics[width=\textwidth]{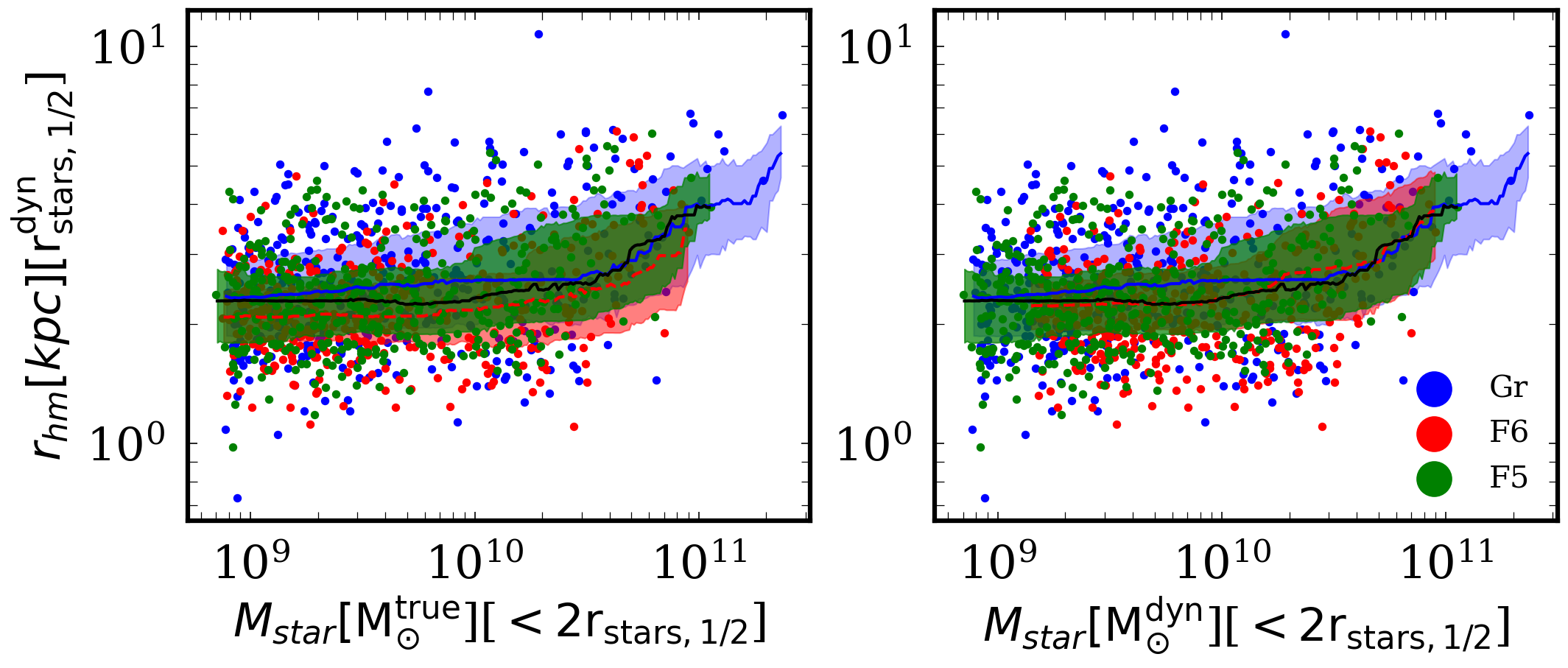}
\centering
\caption{Mass-size relation for the selected haloes, for the GR run (blue), F6 (red) and F5 (green). In solid lines, the moving median. The shaded areas enclose the 25th and 75th quarterlies. Right panel,in terms of the stellar true mass. Left panel, considering the dynamical stellar mass.} 
\label{fig:2A}
\end{figure*}

In Figure \ref{fig:2A} we show the mass-size relation for the selected haloes. The  trend are also similar to the original  {\sc Illustris-TNG} results (see Figure 4 of \citet{Pillepich2017}). Haloes with greater stellar mass have larger sizes (larger stellar half mass radius, $\mathrm{r_{hm}}$). This trend is present independently of  fifth force effects. 

\subsection{Ellipsoid parameters and morphology correlation}

\begin{figure*}
\centering
\includegraphics[width=0.455\textwidth]{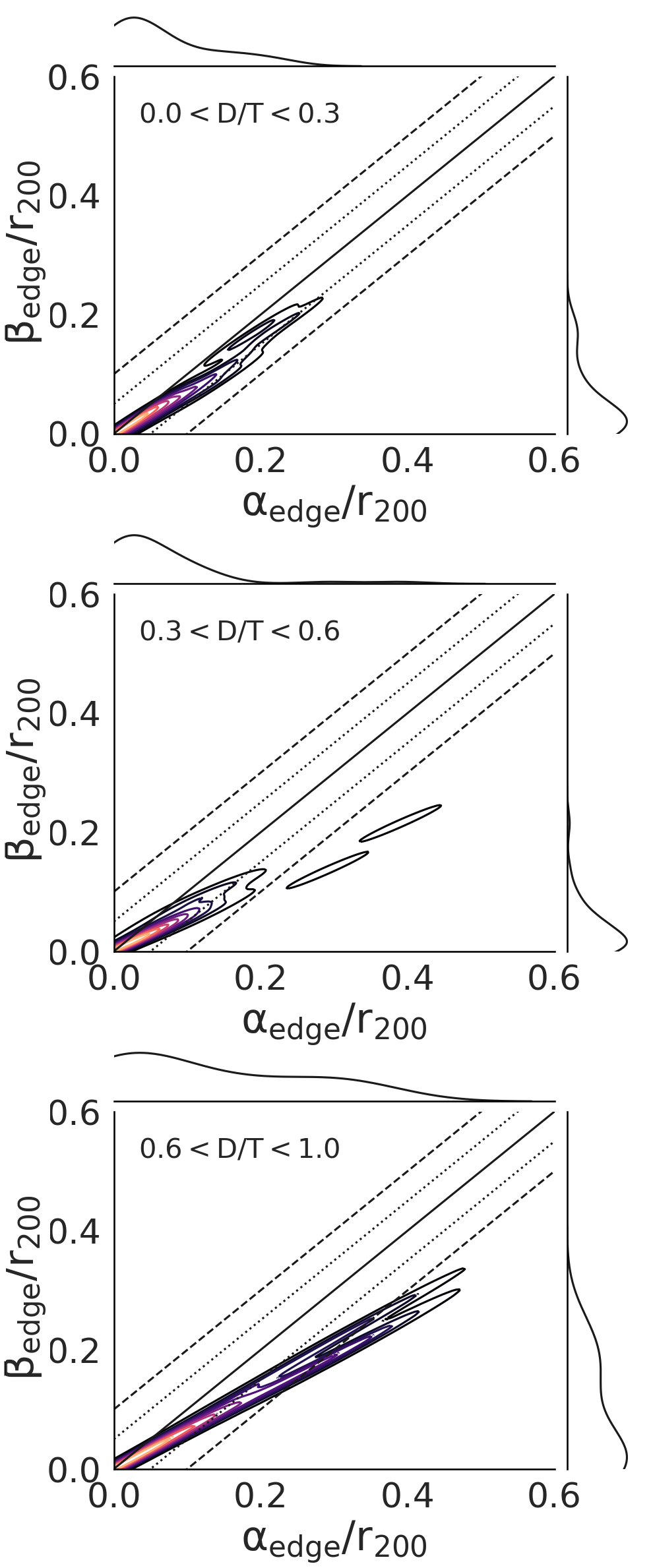}
\includegraphics[width=0.45\textwidth]{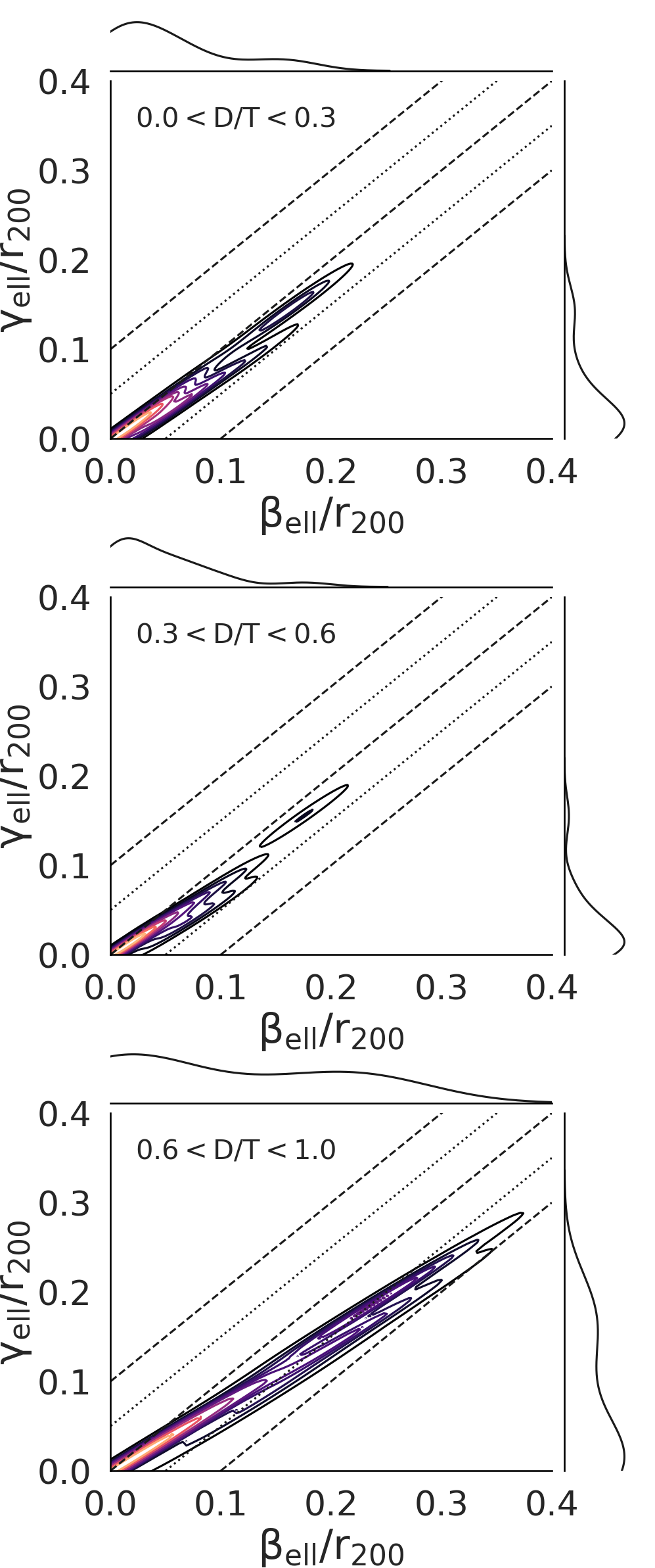}
\caption{The histogram distribution of the ellipsoid (2D) parameters of the ellipses: $\mathrm{\alpha _{edge}/r_{200}}$ vs $\mathrm{\beta_{edge}/r_{200}}$ (left panels). For the ellipsoids (3D), we took $\mathrm{\beta _{ell}/r_{200}}$ vs $\mathrm{\gamma _{ell}/r_{200}}$ (right panels). Each row represents a $\mathrm{D/T}$ bin and their deviation from equality. We define the screening regions as the radii with field values less than $\mathrm{\left |f_{R}/f_{R0}  \right |} = 10^{-2}$. For disc galaxies, the parameters $\beta _{edge}$ and $\gamma _{ell}$ were smaller that the parameters alongside the stellar disc frame, i.e. $\alpha _{edge}$, $\alpha _{ell}$ and $\alpha _{ell}$. See Table \ref{tab:tab_parameter}.}
\label{fig:3A}
\end{figure*}

In Figure \ref{fig:3A}, we plot the 2D Histograms of the shape parameters of the edge-on ellipses ($\mathrm{\alpha _{edge}/r_{200}}$ vs $\mathrm{\beta _{edge}/r_{200}}$) in terms of $\mathrm{D/T}$ (left panels). For the 3D ellipsoids, we show the parameters ($\mathrm{\beta _{ell}/r_{200}}$ vs $\mathrm{\gamma _{ell}/r_{200}}$) (right panels). For larger values of $\mathrm{D/T}$, the spherical screening region breaks up. A more well-defined stellar galaxy disc can be a potential indicator of the morphology of an screening region.

%%%%%%%%%%%%%%%%%%%%%%%%%%%%%%%%%%%%%%%%%%%%%%%%%%

% Don't change these lines
\bsp	% typesetting comment
\label{lastpage}
\end{document}